\documentclass[sigconf]{acmart}

\usepackage{quoting}
\usepackage{color}
\usepackage{xspace}
\usepackage{xcolor}
\usepackage[tight,footnotesize]{subfigure}
\usepackage{multirow}
\usepackage{tikz}
\usepackage{caption}

\definecolor{darkblue}{RGB}{0, 130, 185}
\definecolor{shallowblue}{RGB}{61, 223, 240}
\definecolor{darkyellow}{RGB}{183, 141, 36}
\definecolor{shallowyellow}{RGB}{252, 229, 80}
\definecolor{mygrey}{RGB}{204, 204, 204}

\newcommand{\uline}[1]{\underline{#1}}
\newcommand{\hide}[1]{}
\newcommand{\bluedot}[1]{{\protect\tikz\protect\draw[darkblue,fill=darkblue] (0,0) circle (.5ex); #1}}
\newcommand{\yellowdot}[1]{{\protect\tikz\protect\draw[darkyellow,fill=darkyellow] (0,0) circle (.5ex); #1}}
\newcommand{\greydot}[1]{{\protect\tikz\protect\draw[mygrey,fill=mygrey] (0,0) circle (.5ex); #1}}
\newcommand{\sbdot}[1]{{\protect\tikz\protect\draw[shallowblue,fill=shallowblue] (0,0) circle (.5ex); #1}}
\newcommand{\sydot}[1]{{\protect\tikz\protect\draw[shallowyellow,fill=shallowyellow] (0,0) circle (.5ex); #1}}

\newcommand{\note}[2]{\xspace{\color{#1} [#2]}\xspace}
\newcommand{\todo}[1]{\note{blue}{TO DO: #1}}
\newcommand{\jiao}[1]{\note{red}{Jiao: #1}}
\newcommand{\sherry}[1]{{\note{blue}{Sherry: #1}}}
\newcommand{\vic}[1]{\textcolor{magenta}{(Vic: #1)}}

\newcommand{\diyi}[1]{\note{blue}{diyi: #1}}

\newcommand{\sysname}{GreetA\xspace}
\newcommand{\veryf}[1]{\tikz\draw[red,fill=red] (0,0) circle (.5ex)}

\newcommand{\mkcleanall}{
	\renewcommand{\jiao}[1]{}
	\renewcommand{\sherry}[1]{}
	\renewcommand{\vic}[1]{}
	\renewcommand{\todo}[1]{}
	\renewcommand{\diyi}[1]{}
}
\newcommand{\change}[1]{\textcolor{black}{#1}}
\definecolor{lightblue}{RGB}{53, 122, 191}
\definecolor{lightorange}{RGB}{226, 134, 49}

\mkcleanall

\AtBeginDocument{%
  \providecommand\BibTeX{{%
    \normalfont B\kern-0.5em{\scshape i\kern-0.25em b}\kern-0.8em\TeX}}}

\setcopyright{acmcopyright}

\copyrightyear{2022} 
\acmYear{2022} 
\setcopyright{rightsretained} 
\acmConference[CHI '22]{CHI Conference on Human Factors in Computing Systems}{April 29-May 5, 2022}{New Orleans, LA, USA}
\acmBooktitle{CHI Conference on Human Factors in Computing Systems (CHI '22), April 29-May 5, 2022, New Orleans, LA, USA}
\acmDOI{10.1145/3491102.3502114}
\acmISBN{978-1-4503-9157-3/22/04}

\begin{document}

\title[Pretty Princess vs. Successful Leader]{\emph{Pretty Princess vs. Successful Leader}: Gender Roles in \\ Greeting Card Messages}

\author{Jiao Sun}
\email{jiaosun@usc.edu}
\affiliation{%
  \institution{University of Southern California}
  \country{USA}
}

\author{Tongshuang Wu}
\email{wtshuang@cs.washington.edu}
\affiliation{%
  \institution{University of Washington}
  \country{USA}
  }

\author{Yue Jiang}
\email{yuejiang@mpi-inf.mpg.de}
\affiliation{%
  \institution{Max Planck Institute for Informatics}
  \country{Germany}
}

\author{Ronil Awalegaonkar}
\email{ronil.awale@gmail.com}
\affiliation{%
 \institution{Latin School of Chicago}
 \country{USA}
 }

\author{Xi Victoria Lin}
\email{victorialin@fb.com}
\affiliation{%
  \institution{Meta AI}
  \country{USA}
  }

\author{Diyi Yang}
\email{diyi.yang@cc.gatech.edu}
\affiliation{%
  \institution{Georgia Institute of Technology} 
  \country{USA}
  }

\renewcommand{\shortauthors}{Jiao Sun et al.}

\begin{abstract}
 	People write personalized greeting cards 
on various occasions. While prior 
work has studied gender roles in greeting card messages, 
systematic analysis at scale and 
tools for raising the awareness of gender stereotyping 
remain under-investigated. 
To this end, we collect a 
large greeting card message corpus 
covering three different occasions (birthday, Valentine's Day and wedding) from three sources (exemplars from greeting message websites, real-life greetings from social media and language model generated ones). 
We uncover a wide range of gender stereotypes in this corpus via topic modeling, odds ratio and Word Embedding Association Test (WEAT). 
We further 
conduct a survey to understand people's perception of gender roles in messages from this corpus and if gender stereotyping is a concern. 
The results show that people want to be aware of 
gender roles in the messages, but 
remain unconcerned unless the perceived gender roles conflict with the recipient's true personality. In response, we developed GreetA, an interactive visualization and writing assistant tool to visualize 
fine-grained topics in greeting card messages drafted by the users and the associated gender perception scores, but without suggesting text changes as an intervention. 
\end{abstract}





\maketitle

\section{Introduction}
\label{section:intro}
\begin{quoting}[indentfirst=true]
\emph{Now, the first guests will be arriving in a few minutes, and they are going to find you perfectly behaved, sweet, charming, innocent, attentive, delightful in every way. I particularly wish for that, Lyra, do you understand me?}
\\
\hspace*{\fill}—— \emph{Mrs. Coulter, ``His Dark Materials''}
\end{quoting}

Gender stereotyping 
creates widely accepted biases about certain characteristics of a gender group 
and perpetuates the notion that gender-associated behaviors are binary. 

In many cultures, women are valued by physical attractiveness while men are valued by professional success~\cite{expectation}. Personality-wise, women are supposed to be nurturing and avoid dominance, while men are supposed to be agentic and avoid weakness~\cite{10.3389/fpsyg.2018.01086}. 
Such diminishing, and sometimes negative 
conceptions of 
gender groups are one of the greatest barriers for 
equality and need to be tackled wherever they appear~\cite{gsinads}.

However, gender stereotypes are deeply rooted in society and 
arise in all types of media and social interactions
~\cite{10.3389/fpsyg.2018.01086,gsinads,murphy1994greeting}. 
For instance, advertisement, television and movies 
have all been shown to be glutted with damaging portraits of gender~\cite{gsinads,Steinhagen2010,10.3389/fpsyg.2018.02435,wathinggender} before governments 
start to ban such content~\cite{adsban}.
In this work, we focus on greeting cards messages, a media type that is under-investigated in this context and is exchanged among billions of people 
during holidays and special occasions to express affection, gratitude, sympathy, or other sentiments~\cite{gcindustry}. 

\hide{\begin{figure}
\centering
\begin{minipage}{.5\textwidth}
  \centering
  \includegraphics[width=.83\linewidth]{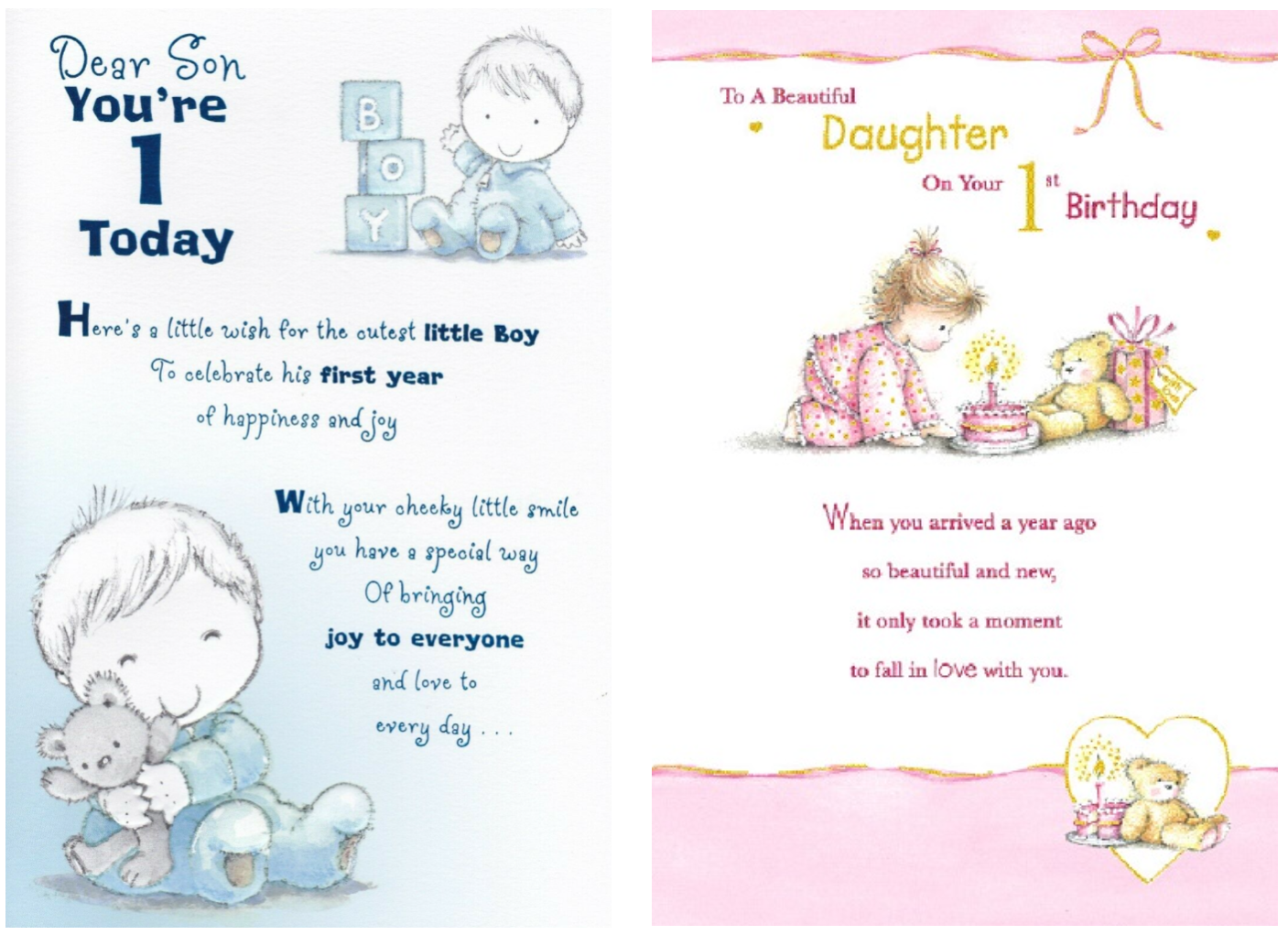}
  \captionof{figure}{}
  \label{fig:test1}
\end{minipage}%
\begin{minipage}{.5\textwidth}
  \centering
  \includegraphics[width=.9\linewidth]{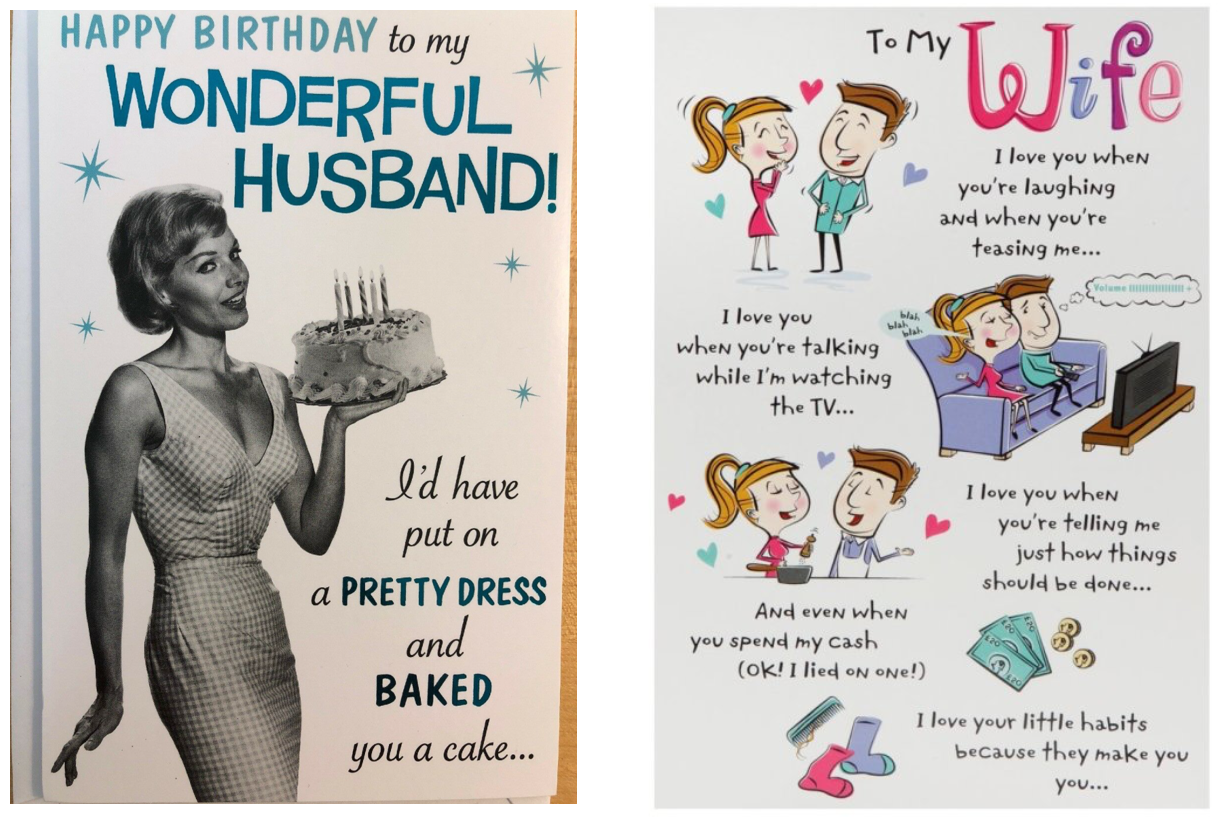}
  \captionof{figure}{}
  \label{fig:test2}
\end{minipage}
\end{figure}

\begin{figure}
\centering
\includegraphics[width=.6\linewidth]{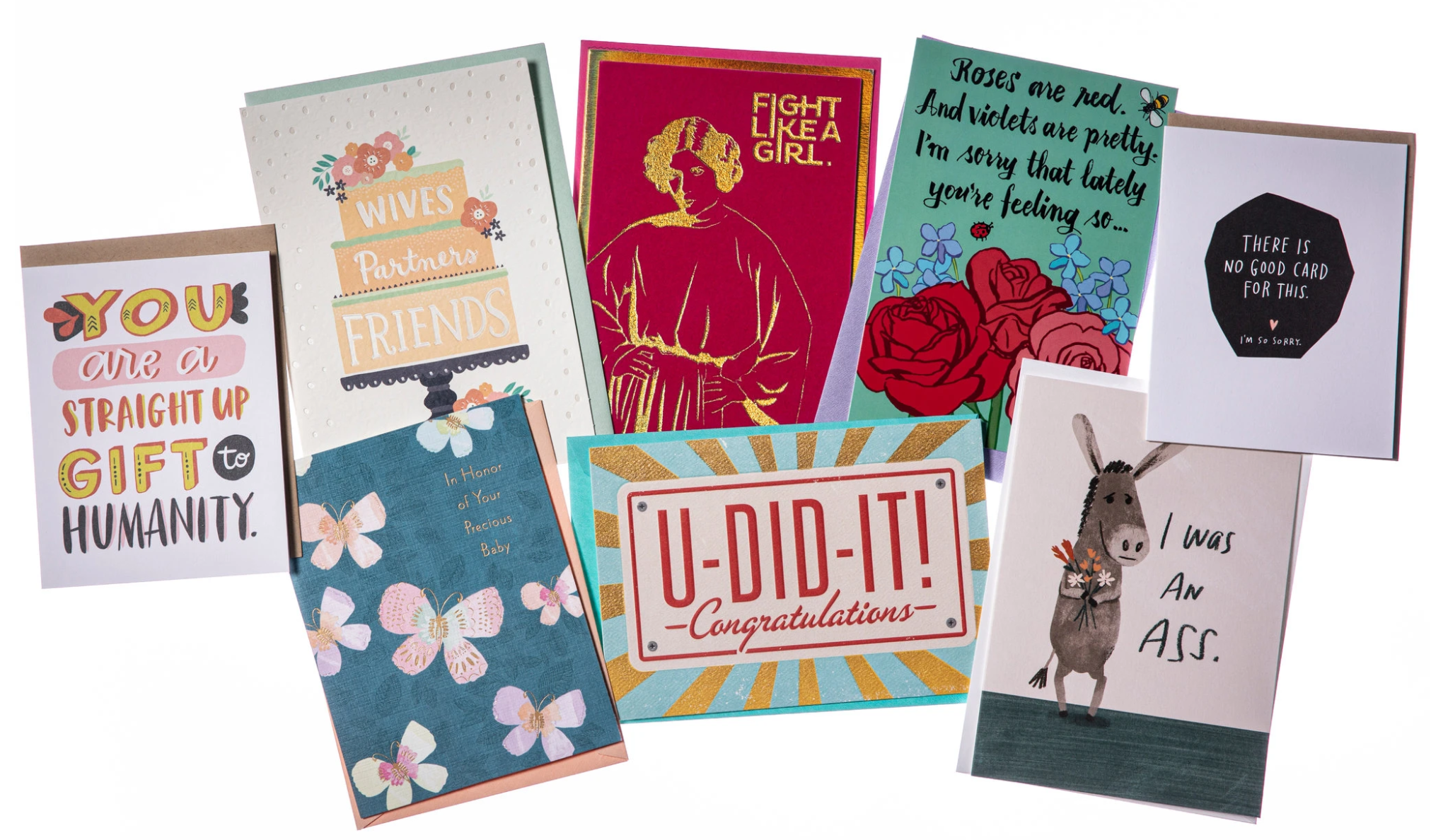}
  \captionof{figure}{}
  \label{fig:assorted_cards}
\end{figure}
}

Greeting cards messages form a compelling source for studying gender stereotypes as they contain both descriptive components (sender's perception of the receiver) and prescriptive components (sender's expectations of the receiver). It is easy to obtain gender information from greeting card messages. People reveal the receiver's identity and gender, sometimes the sender, and the receiver's relationship to the sender in the messages. Greeting card messages significantly impact our everyday life. Since people often share greeting card messages out of the goodwill, gender stereotypes may be enforced without being noticed. We show an example where the greeting for a female recipient is about appearance while for a male recipient is about leadership for their birthdays in Table~\ref{table:example}. Thus, it is important to help people prevent unconscious gender stereotypes. Besides, greeting card messages can be used on various social occasions (e.g., birthdays, weddings, etc.), which offers more opportunities to observe and analyze the gender stereotypes in different settings. This research seeks to answer the following research questions:

\begin{table}[t]
\centering
\begin{tabular}{@{ }p{1.5cm}@{ }  @{ }p{6.1cm}@{ }}
\toprule \textbf{Recipient}& \ \textbf{Greeting Card Message} \\ \midrule
Brother ({\color{lightblue}M})  &  Having a brother who cares like a mother is scarce; \ you have exhibited great {\color{lightblue}leadership} qualities which makes you a successful father, husband, and sibling. Happy Birthday to you. \\
\midrule
Niece ({\color{lightorange}{F}})  & My {\color{lightorange}beautiful} niece is growing a year older that makes me one of the happiest uncles in the world. Have a grand birthday sweetheart; you will always be my {\color{lightorange}princess}. \\
\bottomrule
\end{tabular}
\caption{Examples of gender roles in greeting card messages for birthday from OccasionMessage~\cite{occasionmessages}.}
\vspace{-0.5cm}
\label{table:example}
\end{table}

\begin{itemize}
	\item [\textbf{RQ1}] Would there be gender stereotypes in greeting card messages? To \change{women} lean towards their beauties and household, and to \change{men} lean towards their work achievement?
	\item [\textbf{RQ2}] If RQ1 stands, would the gender association of topics in greeting card messages relate to recipients' age?
	\item [\textbf{RQ3}] Do people want to be informed of the gender role \change{(masculinity and femininity for our study)} in their greeting card messages?
	\item [\textbf{RQ4}] If RQ3 stands, how can we help increase the gender role awareness to people? What features do they expect?
\end{itemize}

To quantify these problems, we scraped over 18,000 greeting card messages from eight popular websites (including HallMark~\cite{HallMark} and American Greetings~\cite{AmericanGreetings}) to form our \emph{template dataset}, which covers three different scenarios: Birthday, Valentine's Day and Wedding.
We extract topics from greeting card messages, further select the ones that have gender associations, and do fine-grained analysis for different age groups. Besides templates, we are also interested in studying greeting messages written by people. We approximate greetings written by people using those generated by the state-of-the-art language model GPT-2~\cite{GPT-2}, which are trained over an ocean of text on the web. Our analysis shows that greeting card messages to \change{women} lean towards their appearance and household and to \change{men} lean towards work achievement, and greetings to the elderly have less association with gender compared to other age groups. We obtained similar results on greeting messages generated by language models. 


\hide{While many work on gender bias focus on providing writing suggestions/helping people to mitigate the bias, such intervention may not be desirable in our case, because sometimes the distinction is justified depending on the recipients' self-identified gender. }
To understand how people perceive the potential gender association of topics in their own messages, we designed a pilot survey
and collected feedback from twenty users with a diverse \change{ethnical background} spanning from their 20s to 50s. 
According to the survey, most people would like to be informed of the gender role and avoid potential gender stereotypes in their greeting card messages, but do not want the machine to intervene much or modify their messages.
Therefore, we 
develop \sysname (\textbf{Greet}ing with Gender Role \textbf{A}wareness), a visualization tool that helps users write greeting card messages with the gender role awareness. 

We further verify the effectiveness of \sysname by conducting comprehensive qualitative and quantitative user studies.
Our qualitative study indicates that most of participants agree that \sysname is useful for writing greetings, is easy to learn and use, and they would like to use it in the future. In addition, three contrast surveys in the quantitative study show that \sysname helps people increase the gender role awareness when they write greeting card messages.
These user studies explore people's perceptions of gender association in greeting card messages and show that \sysname effectively assists people in being more aware of gender association at scale.

In summary, our work demonstrates the following contributions: 
\sherry{Strategically, when reviewers ask about framing/scoping question and/or want us to think about ethical limitations, it's usually good to also highlight them in the intro.}
\begin{itemize}
    \item  \emph{\textbf{Dataset.}} We \change{collect} a large-scale greeting template dataset with gender information, with over eighteen thousand messages and six greeting scenarios. ~\footnote{\change{We discarded the ``congratulations'', ``baby shower'' and ``sympathy'' scenarios for analysis because of their small size. In result, we kept and used ``Birthday'', ``Valentine's Day'' and ``Wedding'' scenarios for analysis throughout the paper.}} We also \change{contribute} an AI-generated (i.e., GPT-2) greeting message corpus \change{together with a birthday greeting Tweets dataset}  to facilitate future research on greeting card messages.
    
	\item \emph{\textbf{Analysis.}} To the best of our knowledge, we are the first to analyze gender roles in large-scale greeting card messages quantitatively using statistical NLP tools. We further analyze greeting messages among different age groups and scenarios. 
	
	\item \emph{\textbf{System.}} Based on the analysis of users' feedback and requirements, we design and build a supportive visualization tool, named \emph{\sysname}, to help users write greeting card messages with gender role awareness. 
	
	\item \emph{\textbf{Evaluation.}} To evaluate \sysname, we conduct both a qualitative user study and a quantitative user study. We found that \sysname is easy to use and helps users increase gender awareness when writing greeting card messages.
	
\end{itemize}

\section{Related Work}

\paragraph{Gender Related Research. } In the HCI community, ~\citet{Burtscher2020ButWW} provide a staring point for HCI researchers to explore questions and issues around gender. ~\citet{Stumpf2020GenderInclusiveHR} give a conceptual review and provide some evidence for the impact of gender in thinking and behavior which underlines HCI research and design, and ~\citet{Schlesinger2017IntersectionalHE} introduce a framework for engaging with complexity of users' multi-faceted identities including demographic information. Meanwhile, researchers in the social computing community have been studying the gender role in various social settings. For instance, prior work found that women contribute far less frequently than men in the Question and Answer sites and analyzed how the community cultures might be impacting men and women differently\change{~\cite{gender_qa, gender_representation}}. ~\citet{freelance} showed that the perceived gender is significantly correlated with worker evaluations. In addition to the inequalities caused by the external factors, ~\citet{foong2018women} find that the female workers set a self-determined hourly wage lower than males' in an online labor marketplace. Males also express higher negativity and lower desire for social support when they face mental illness \change{\cite{McKenzie2018MasculinitySC}}. Similarly, ~\citet{McGregor2017PersonalizationGA} find that male candidates may see more and female candidates see less strategic benefits in personalizing campaign politics on social media. More relatable to our research, ~\citet{Reifman2020HappyanniversaryGA} find gender and age differences when expressing vulnerable emotions such as love in anniversary greetings delivered on Twitter.

Gender-related research has also been an emerging topic for Artificial Intelligence (AI). In the computer vision field, ~\citet{computers} found that facial analysis technologies performed consistently worse on transgender individuals and were unable to classify non-binary genders.  \change{In the natural language processing field, researchers measured the gender gap in various downstream tasks (e.g., authorship and citations~\cite{Mohammad2020GenderGI}), raising ethical considerations to use gender as a variable in NLP. In response, researchers have been addressing the gender awareness and making efforts to alleviate the gender bias in models (e.g.,machine translation field~\cite{stanovsky-etal-2019-evaluating, saunders-byrne-2020-reducing}}. Besides, researchers have discovered gender bias in pretrained models (word embeddings)~\cite{Zhao2019GenderBI}, data~\cite{Hovy2016TheSI}, and algorithms themselves~\cite{Basta2019EvaluatingTU}. ~\citet{Cryan2020DetectingGS} compared the lexicon and supervised methods for detecting gender stereotypes.  
Compared to prior work that tries to mitigate gender bias~\cite{Wang2019BalancedDA}, our work \emph{provides gender role awareness} and \emph{prevent potential gender stereotypes}. 

\paragraph{Gender Role in Greetings Card Messages. } Psychology studies showed that greeting card messages are an important source of the self-concept establishment \change{\cite{murphy1994greeting}}, which has a long-term impact on individuals~\cite{McCall2007TheMO}. ~\citet{West2009DoingGD} illustrated that greeting card communication reflects the highly gendered division in the U.S. culture. Few prior research has been focused on analyzing greeting card messages due to limited data. ~\citet{murphy1994greeting} analyzed 180 Hallmark children's cards in a real store and found out that girls and boys are perceived as being different in terms of interests, activity levels, and characteristics. They further identified children's greeting cards as an arena of clear gender-associated messages. To facilitate further exploration, we collect a large dataset from popular websites of greeting message suggestion websites along with AI-generated greeting messages. We analyze the dataset quantitatively to discover and understand the gender association of topics in greeting card messages. Although the topic of greeting cards and gender has been visited in a series of gender research literature ~\cite{Keith2009HailingGT}, to our best knowledge, we are the first to analyze this issue quantitatively using statistical NLP tools. 

\paragraph{Quantify Gender Bias. }
Previous research has explored how to measure gender bias. \change{Researchers~\cite{Zhou2019ExaminingGB, May2019OnMS} have been widely using Word Embedding Association Test (WEAT) scores~\cite{WEAT} to quantify gender bias.} The WEAT score links bias in word embeddings to human bias. It compares two sets of target words (e.g., art and science words) and a pair of opposing attribute words (e.g., female and male names). Besides, the WEAT score measures the association strength between the target words group and the attribute word group using vector similarities in the word embedding. In this paper, we use the WEAT score to qualify gender bias in greeting messages.

\paragraph{Text Generation with Artificial Intelligence (AI)} With the astounding growth of AI, people have been using it for dialogue systems~\cite{Sordoni2015ANN}, summarization~\cite{Nallapati2016AbstractiveTS}, story generation~\cite{Roemmele2016WritingSW}, etc. GPT-2, openAI's publicly available language model, and BERT~\cite{Devlin2019BERTPO} are the state-of-the-art AI models for text generation. ~\citet{Basta2019EvaluatingTU} find that GPT-2 performs better than BERT. 
In our work, we use GPT-2 to generate greeting card messages given some prompts as input, reflecting real greetings it has trained on. We found that GPT-2 generated greeting card messages also have similar gender associations to the template dataset.
\section{Analysis}
\label{section:analysis}

We aim to explore the following two research questions:

\begin{itemize}
	\item [\textbf{RQ1}] \emph{Would greeting card messages to females lean towards their beauties and household, and to males lean towards their work achievement?} \change{Prior work ~\cite{murphy1994greeting} has shown that gender signals exist in greeting card messages and make the rejection of gender stereotypes difficult.}    
	We are interested in understanding whether such distinctions hold across various messages from multiple scenarios.
	\item [\textbf{RQ2}] \emph{If RQ1 stands, would the gender association of topics in greeting card messages relate to recipients' age?}
	\citet{murphy1994greeting} focused on messages whose recipients are children; However, messages sent to adults are equally important, as the accumulation of gender signals makes the stereotype difficult to overcome and has long-lasting effects in various aspects including career, mental health, etc~\cite{McCall2007TheMO}. We are interested in understanding whether the distinctions are stably maintained across age groups, or whether it diminishes or strengthens.
\end{itemize}

To answer the two questions, we collect and analyze messages from \change{three} data sources. 
First, we collect a large-scale greeting card message corpus under six scenarios using templates from greeting message suggestion webpages (we refer to it as \emph{Template} Dataset).
As these examples demonstrating ``ideal'' sample messages, we believe they reflect people's expectations on greeting message writing.
Second, we also generate artificial greeting messages from large natural language models (referred to as \emph{GPT-2} Dataset).
These models are usually trained on large text corpus crawled from the Internet. Therefore, their generation-- if high quality-- can reflect real messages users write and post. 
\change{Third, we manually collect 500 tweets from real users on Twitter to analyze how users write greeting card messages naturally in the real world.}
Below, we first introduce our data collection strategy and then illustrate the analysis process.

\subsection{Data}
\label{section:data}

\renewcommand{\arraystretch}{1.2}
\begin{table}[]
\small
\begin{tabular}{p{0.25\linewidth} p{0.68\linewidth}}
\toprule
\textbf{Group} & \textbf{Gender Indicators} \\
\midrule
General \texttt{female} & daughter, hers, lady, grandma, grandmother, female, aunt, wife, sis, niece, mother, she,  girl, her, granny, granddaughter, girlfriend, woman, mom, sister \\
\hline
General \texttt{male}   & dude, godfather, grandson, stepbrother, boy, sir, he, uncle, man, male, soninlaw, boyfriend, brother, grandpa, him, nephew, son, papa, exboyfriend, granddad, husband, stepson, dad, fatherinlaw, daddy, stepdad, father, grandfather, bro, his  \\ \hline
\texttt{Mother} variations  & mother, mom, mama, mommy, mum, mumsy, mamacita, ma, mam, mammy \\ \hline
\texttt{Father} variations & father, dad, dada, daddy, baba, papa, pappa, papasita, pa, pap, pop\\ \hline
\texttt{Grandmother} variations   & grandmother, grandma, grandmom, grandmama, grama, granny, gran, nanny, nan, mammaw, meemaw, grammy \\ \hline
\texttt{Grandfather} variations   & grandfather, grandpa, gramp, gramps, grampa, grandpap, granda, grampy, granddad, grandad, granddaddy, grandpappy, pop, pap, pappy, pawpaw \\
\bottomrule
\end{tabular}
\caption{
We used a list of recipient indicators (i.e., ``attributes''~\cite{WEAT}) to 1) prompt the message generation in GPT-2 dataset collection (Section~\ref{section:data}), and 2) distinguish female- and male- associated messages in the template dataset. The General \texttt{female} and \texttt{male} indicators are from ~\citet{WEAT} and addresses from the template dataset, used to generate general wishes. Other variations are terms of endearment from their corresponding Wikipedia pages and used to generate wishes for different age groups.}
 
\label{table:gender_indicators}
\end{table}

\paragraph{Template Dataset}
We crawled eight websites~\cite{HallMark, AmericanGreetings, bestcardmessages, greeting-card-messages, wishesmsg, holidaycardapp, occasionmessages, 143greetings}. \change{Our collected greeting card messages are all publicly available. We only use collected data for personal research purposes, which is compliant with all platforms' terms of service.}
Among them, Hallmark and AmericanGreetings are the two largest greeting card producers globally, and the greeting messages released on their web pages cover the widely-held social values regarding various topics \cite{murphy1994greeting}. 
We use other websites as supplements that enlarge the variety of our datasets. In total, we collected 18,559 messages of 6  common scenarios across all websites, indicating the popularity of these scenarios, \change{among which we choose 3 for our analysis.}
We further split these messages based on their gender association.
We assume that the association is implied by the recipient's gender mentioned in the message: If a message mentions an indicator in the general female group in Table~\ref{table:gender_indicators} (or variations of \texttt{mother} or \texttt{grandmother}), we categorize it as a \emph{female-associated message.} 
\change{We mark the messages with unknown recipients as \emph{neutral} and will be using them to study how people write greeting card messages to recipients without gender-specific indicators (e.g., ``sincere birthday messages'' in HallMark and ``for Coworkers'' in Americangreetings)}.
\sherry{I like this!}
The resulting distribution is in Table~\ref{table:statistics}. \change{Note that we do not consider or differentiate greeting card messages based on senders' gender, meaning that our collected dataset may include greeting card messages among the same gender.}

\renewcommand{\arraystretch}{1.0}
\begin{table}[]
 \setlength{\tabcolsep}{10pt}
\resizebox{0.48\textwidth}{!}{
\begin{tabular}{l l l}
\toprule

\textbf{Scenario} & \textbf{Template Dataset}  & \textbf{GPT-2 Dataset}\\ \midrule
Birthday        &  13,338 ({\color{darkyellow}4138}/{\color{darkblue}4208}/{\color{gray}4992})  &  40,200 ({\color{darkyellow}19,600}/{\color{darkblue}20,600}/{\color{gray}0})     \\ 
Valentine's Day &  2,496 ({\color{darkyellow}610}/{\color{darkblue}654}/{\color{gray}1232})  &  28,200     ({\color{darkyellow}13,600}/{\color{darkblue}14,600}/{\color{gray}0}) \\ 
Wedding         &  1,360 ({\color{darkyellow}121}/{\color{darkblue}122}/{\color{gray}1117})  &  28,200     ({\color{darkyellow}13,600}/{\color{darkblue}14,600}/{\color{gray}0})  \\ 
 \bottomrule
\end{tabular}
}
 \caption{Statistics showing the number of greeting card messages we collected in the Template and the GPT-2 datasets. The numbers in parentheses indicate the number of messages for {\color{darkyellow}female}, {\color{darkblue}male}, and {\color{gray}neutral} messages respectively. In the GPT-2 dataset, for each scenario, we generated 200 messages for each recipient described in Table~\ref{table:gender_indicators}. 
 }
\label{table:statistics}
\end{table}

\paragraph{GPT-2. }
\change{The rapid development of AI has enabled text generation models to generate fluent content like humans. Therefore, we are also interested in studying whether greeting card messages written by AI will have similar patterns to human-written greeting card messages. In this work, we use a state-of-the-art language model GPT-2 as a black box tool to generate greeting card messages.} 
Following similar approaches from prior work (e.g.,~\citet{vig2020causal} studying gender bias), we feed the GPT-2 model with different prompts (i.e., partial keywords or phrases like ``Happy birthday mom!''), and collect the continuation generated by the model. The intuition is that, if the model is trained on gender-distinctive messages, prompts with clear gender indications will trigger it to generate messages that have reflect such distinctions.
We configured the model to use 
top-p~\cite{topp} ($p=0.1$) sampling, and constrained the length of the generated sentence to be 200 characters.
Such setting maximizes
the chance of getting non-repetitive and natural greeting messages based on our experiment.
We designed the prompts in the following ways. \emph{For general wishes, }we use two kinds of prompts:
\begin{itemize}
\item  \texttt{{``[scenario prefix] [female/male indicator]!''}}, with the \texttt{scenario prefix} being appropriate wish sentence for one of the six scenarios from Table~\ref{table:statistics} (e.g., ``Happy birthday'' for \texttt{birthday}, ``Congratulations on getting married'' for \texttt{wedding}), and the female/male attributes and variations from Table~\ref{table:gender_indicators};
\item \texttt{``[scenario prefix] [female/male name]!''}, to cope with the fact that people often refer to their loved ones by name. We adopted common female/male names from ~\citet{checklist}.
\end{itemize}

\emph{We further generate messages for \emph{babies}, \emph{parents} and \emph{grandparents}} to check if the gender association of topics varies for different age groups. 

\begin{itemize}
\item  \change{For \emph{babies}, we use \texttt{``[scenario prefix] my little baby [girl/boy] [female/male name]!''} as prompts to generate birthday messages.} 
\item For \emph{parents} and \emph{grandparents}, we use \texttt{``[scenario prefix] [corresponding terms of endearment]!''} as prompts. We show terms of endearments for parents and grandparents age groups in Table~\ref{table:gender_indicators}. 
\end{itemize}

For each scenario and gender indicator, we generated two hundred messages.
As shown in Table~\ref{table:statistics}, GPT-2 generates sufficient messages so that we can generate enough data for all age groups. 
Note that we do not generate greeting messages under \emph{Valentine's Day} and \emph{wedding} scenarios for baby age group, which leads to the statistical difference in Table~\ref{table:statistics}.

\change{\paragraph{Twitter Data.} \emph{Template dataset} represents what people regard as ideal messages, and \emph{GPT-2} generated greeting card messages represent how model would write after seeing abundant online text. But either case might not represent well how people actually write greeting card messages in the real life, which is usually private and not accessible. Although people have been using social media to greet others, it is hard to acquire recipients' gender automatically because of the difficulty of extracting the recipient from the Tweet content and getting gender identity information from another linked account (i.e., recipient account). To address this issue, we design three criteria to collect Tweets of birthday greetings where we can identify recipients' gender: 1) the content of Tweets have to be birthday greetings to another recipient; 2) senders need to explicitly mention the recipient (by using \texttt{@}) in their posts, and 3) recipients need to self-identify their gender by explicit means (e.g., mention gender pronouns) in their public profiles. In our collection process, we search for keyword \emph{``birthday''} and put time range from \emph{2019/01/01} to \emph{2021/03/01}. To ensure the high data quality, three of coauthors go to Twitter and manually collected 500 tweets about birthday greetings that satisfy both our criteria, so that we can acquire recipients' gender identities. 
Among 500 tweets we collected, 263 are for women and 236 are men. }



\subsection{Analysis Methodology}
\label{section:method}

\begin{figure*}
    \centering
    \includegraphics[width=1.8\columnwidth]{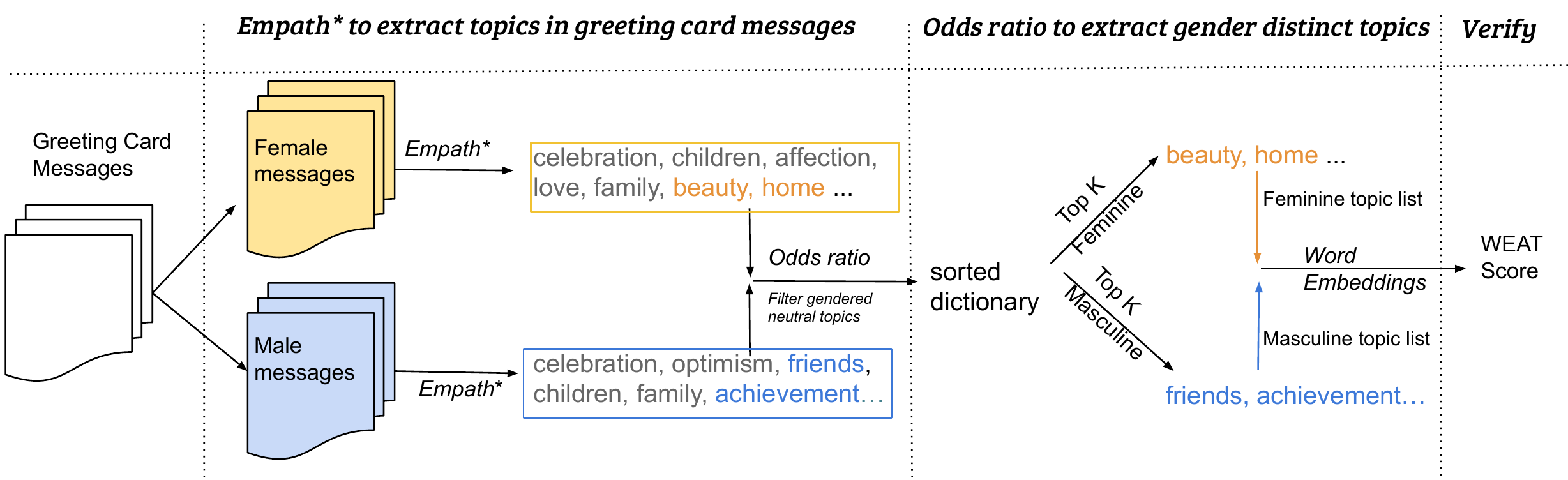}
    \caption{Analysis pipeline. We first apply a modified version of Empath~\cite{Fast2016EmpathUT} (Empath*) to extract topics from greeting card messages. We then apply the odds ratio~\cite{odds-ratio} to extract gender distinct topics. Finally, we quantify and verify the association of topics and gender \change{with WEAT scores}~\cite{WEAT}.}
    \label{fig:pipeline}
\end{figure*}


\begin{figure*}
    \centering
    \includegraphics[width=1.8\columnwidth]{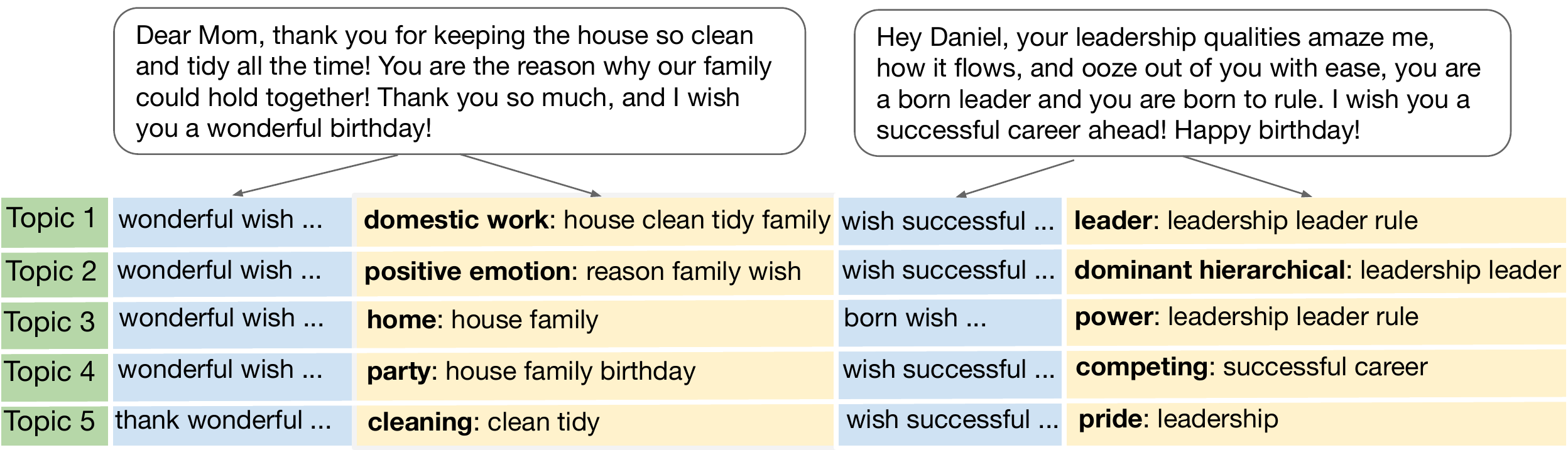}
    \caption{\change{Topics generated by LDA~\cite{LDA} (in blue) severely suffer from repetition and ambiguity issues. While Empath~\cite{Fast2016EmpathUT} (in yellow), used in our pipeline, could accurately generate suggested topics and corresponding words falling into each topic.}}
    \label{fig:lda_vs_empath}
\end{figure*}

\change{In this section, we illustrate our analysis pipeline in Figure~\ref{fig:pipeline} under the binary gender setting. 
It is worth mentioning that our pipeline is generic, and can be easily adapted to analyze any other gender groups by replacing female recipients and male recipients, e.g., recipients with gender indicators and recipients without specific gender indicators. Our analysis pipeline is as follows:} \sherry{good!}

First, we modify and apply Empath~\cite{Fast2016EmpathUT} to extract topics from greeting card messages. 
Then, we apply the odds ratio~\cite{odds-ratio} to extract gender distinct topics. Finally, we quantify and verify the association of topics and gender with the WEAT score~\cite{WEAT}. We describe the detailed analysis steps as below.


\subsubsection{Step 1: Topic modeling} 
\label{subsubsec:split}
\mbox{} \\
While counting word frequencies might be most intuitive for distinguishing female- and male- associated messages, word-level analysis can be too diverging and coarse to reveal higher-level themes of the messages.
Instead, we choose to associate topics with messages and analyze greeting card messages from the topic level.

 To model the topics, we first tried Latent Dirichlet Allocation (LDA)~\cite{LDA}, a classical generative statistical model that extracts topics in an unsupervised manner.
However, Figure~\ref{fig:lda_vs_empath} clearly shows that LDA tends to deliver repetitive topics.
Moreover, with the topics being exact keywords extracted from the documents, they tend to be ambiguous and incomprehensible. 
We instead applied Empath~\cite{Fast2016EmpathUT} to mitigate these problems.
Built on top of neural embeddings and crowdsourcing, Empath provides mappings between 60k common tokens and 200 high-level topics; Accordingly, we can extract its topic by counting the keyword occurrence in a message. For example, keywords \emph{house, clean, tidy, family} point to the topic \textbf{domestic work} in Figure~\ref{fig:lda_vs_empath}). \change{We conduct a small-scale qualitative evaluation of Empath versus LDA on 30 greeting card messages randomly sampled from the template dataset. First, we run Empath and LDA separately for every greeting card message, and get the first 5 topics together with all words in each topic from the model output.  Then, we hide the model name and ask two coauthors to qualitatively choose which topic model outputs better topics with corresponding words under each topic. Two coauthors unanimously choose Empath for all 30 messages, which qualitatively shows the superiority of Empath over LDA.} \diyi{any further details here? What did you show to annotators? The current description looks not very convincing. }
\sherry{Would it be possible to find a non-coauthor to do this? Seems rather biased if we are evaluating our own. As for Diyi's question. I think you can try to just say we give examples as in figure 2.}

\renewcommand{\arraystretch}{1.1}
\begin{table}[]
\resizebox{0.48\textwidth}{!}{
\begin{tabular}{p{0.07\linewidth} p{0.85\linewidth}}
\toprule
\textbf{Topic}     & \textbf{Message}             \\ \midrule
\rotatebox[origin=r]{90}{Competing} & 
    You have {\color{teal} enlightened} our world with your love and smile...\newline
    You will continue to rise in {\color{teal}wisdom} and all the good things of life will come to you...\newline 
    You're filled with {\color{teal}wisdom} and special activities of yours always revealed this...\newline You process both physical and {\color{teal}intellectual} qualities...\\ \midrule
\rotatebox[origin=r]{90}{Crime} & 
    ... We are highly delighted to {\color{teal}witness} today...\newline 
    ... You're renewed by the privilege given to {\color{teal}witness} another year...\newline 
    I already rescheduled all my appointment for today to enable me to {\color{teal}witness} your birthday party...\newline 
    Your cuteness is the most {\color{teal}killing} weapon...\newline
    ... Your smile is my {\color{teal}killing} weapon... \\ 
\bottomrule
\end{tabular}
}
\caption{\change{Examples of suggested top topics generated from Empath~\cite{Fast2016EmpathUT} and original messages, where we highlight words that Empath catches as related to the suggested topic. Most topics extracted and suggested by Empath correctly reflect gist from greeting card messages ({\em e.g.}, \emph{``competing''}), but can also be misled by the frequent occurrence of some keywords and extracts unwanted topics ({\em e.g.}, \emph{``crime''}). We propose Empath$^{*}$ to alleviate the impact of having frequent outlier keywords on extracting topics.}}
\label{table:examples}
\end{table}


Since the associated keywords in Empath are taken out-of-context.{\change{\footnote{\change{The list of topic-keywords mapping from existing work \citet{Fast2016EmpathUT} can be found here \url{https://github.com/Ejhfast/empath-client/blob/master/empath/data/categories.tsv}. Although Empath demonstrates its efficiency in extracting topics, the out-of-context mapping will sometimes lead to unsatisfying topics. We will introduce a critical evaluation of Empath as a limitation in Section~\ref{section:limitation}.}}}} directly counting the occurrence of these words does not always yield reasonable topics.
For example, in Table~\ref{table:examples}, while keywords \emph{``Enlighten'', ``wisdom'', ``intellectual'', ``religious''} correctly reflect the \emph{``competing''} topic, it seems unreasonable to highlight the topic \emph{``crime''}, just because a single keyword, \emph{``witness''} or \emph{``killing''}, is used as metaphors.
To discourage outliers, we filter the topics for each message based on the diversity and frequency of their corresponding keywords:
We only keep a topic for a message, if 
\change{1) more than 5 of its unique keywords under the specific topic occurring in the message}, and 
\change{2) the average occurrence frequency of each keyword is more than 3}.
We refer to the Empath with filtering as \emph{Empath*}. \change{Note that we choose the thresholds based on running multiple experiments and manually check if there are strange topics as outliers from the output, as there is no ground truth in such an unsupervised setting. Although the empirical numbers may not apply to other scenarios, we advocate researchers to consider using \emph{number of unique keywords under one topic} and \emph{average occurrence frequency of each keyword} as filters adapting to the new setting.}



By applying Empath* to female and male greeting card messages separately, we get two topic dictionaries, with both topic names as keys and their corresponding keyword occurrence frequencies. \change{As in Figure~\ref{fig:pipeline}, \emph{``celebration''} is the most frequent topic in messages for both men and women.}

\subsubsection{Step 2: Use the odds ratio to extract gender distinct topics.} 
\mbox{} \\
With the topic list, we apply odds ratio (OR)~\cite{odds-ratio} to understand if the topics indeed associate with recipients' genders.
~\citet{grantpeer} use OR to analyze the gender differences in grant award procedures.
In our case, OR quantifies the strength of gender association between the two topic lists. 
Intuitively, OR equal to one means that the odds of having one topic in messages to \change{women} is the same as in those to \change{men}. 
If OR is greater than one, the topic is more likely to occur in messages to males, and vice versa. 
Sorting the topics by its computed OR gives us top topics associated with messages to male or female recipients.
To focus our analysis on the most distinctive topics, \change{we filter topics that are under 30\% quantile and keep the same number of topics associated with female/male recipients. Note that we decide the threshold by running multiple sets of experiments after checking the output topics manually. } 
\change{We also calculate the odds ratio gap between the most masculine and most feminine words to quantify the polarity of gender role in greeting card messages.}

\subsubsection{\change{Step 3: Calculate WEAT scores to confirm the distinction.}}
\mbox{} \\
\change{Are the topics we pick indeed associated recipients' genders?} We further validate whether the topics we select are associated with gender attributes in popular neural word embeddings like GloVe~\cite{glove}.
We quantify this with the Word Embedding Association Test (WEAT)~\cite{WEAT}, a popular method for measuring biases in word embeddings. 
Intuitively, WEAT takes a list of tokens that represent a concept (in our case, \emph{keywords} for each \emph{topic}) and verifies whether these tokens have a shorter distance towards female attributes or male attributes (in our case, the \emph{indicators} from Table~\ref{table:gender_indicators}).
We calculate the score on three versions of pretrained GloVe embeddings, including Google News, and Wikipedia, and Gigaword~\cite{Gigaword}.
To better represent words in our corpus, we also fine-tune GloVe on our corpus with Mitten~\cite{mitten}.
We refer to the fine-tuned GloVe embedding on our corpus as GloVe*. 

%

\subsection{Analysis Result}
\begin{table*}[]
\begin{tabular}{l l l l l}
\toprule
\multicolumn{2}{l}{Group}  & Feminine topics                                                                                            & Masculine topics   & Gap                                                                                     \\ \midrule
\multicolumn{2}{l}{T-all}     & \begin{tabular}[c]{@{}l@{}}\yellowdot{} feminine, \greydot{}{} appearance, \sydot{} attractive, \\ \greydot{}{} aggression, \yellowdot{} beauty\end{tabular}          & \begin{tabular}[c]{@{}l@{}}\bluedot{} masculine, \bluedot{} leader, \sbdot{} work, \\  \sbdot{} philosophy, \bluedot{} social media\end{tabular}   & 3.50\\  \hline
\multicolumn{2}{l}{T-babies}  & \begin{tabular}[c]{@{}l@{}}  \yellowdot{} \underline{feminine}, \greydot{}{} childish, \sydot{} affection, \\ \yellowdot{} shape/size, \yellowdot{} friends\end{tabular}            & \begin{tabular}[c]{@{}l@{}}\bluedot{} achievement, \greydot{}{} home, \greydot{}{} cold, \\ \greydot{}{} ancient, \greydot{}{} violence\end{tabular}  & 1.46    \\ \hline
\multicolumn{2}{l}{T-parents} & 
    \begin{tabular}[c]{@{}l@{}} 
        \sydot{} \underline{feminine}, \sydot{} \uline{appearance}, \\ 
        \sydot{} domestic work,  \greydot{}{} \uline{beauty}, \greydot{}{} confusion
    \end{tabular}
    & \begin{tabular}[c]{@{}l@{}}  
        \sbdot{} \underline{masculine}, \greydot{}{} ancient, \greydot{}{} heroic, \\ 
        \greydot{}{} business, \sbdot{} fun
        \end{tabular}   
    & 4.19     
\\ 
\hline
\multicolumn{2}{l}{T-grand}   & \begin{tabular}[c]{@{}l@{}}\yellowdot{} love, \sydot{} affection, \greydot{}{} home, \\ \yellowdot{} family, \sydot{} ancient\end{tabular}                          & \begin{tabular}[c]{@{}l@{}}\change{\sbdot{} healing}, \sbdot{} celebration,\\ \sbdot{} optimism,  \sbdot{} friends, \sbdot{} rural\end{tabular}   & \textbf{0.94}       \\ \toprule
\multicolumn{2}{l}{G-all}     & \begin{tabular}[c]{@{}l@{}}\yellowdot{} feminine, \sydot{} attractive, \sydot{} appearance, \\ \yellowdot{} childish, \sydot{} white collar job\end{tabular}  & \begin{tabular}[c]{@{}l@{}}\bluedot{} masculine, \greydot{}{} leader, \sbdot{} play, \\ \sbdot{} wedding, \greydot{}{} driving\end{tabular}  & 3.13              \\ \hline
\multicolumn{2}{l}{G-babies}  & \begin{tabular}[c]{@{}l@{}} \yellowdot{} \underline{feminine}, \yellowdot{} \underline{childish}, \sydot{} royalty, \\ \greydot{}{} ocean, \greydot{}{} magic\end{tabular}                      & \begin{tabular}[c]{@{}l@{}} \underline{\sbdot{} play}, \greydot{}{} pet, \greydot{}{} technology, \\ \greydot{}{} animal, \greydot{}{} hipster \end{tabular} & 9.05
\\ \hline
\multicolumn{2}{l}{G-parents} & \begin{tabular}[c]{@{}l@{}}\yellowdot{} domestic work, \greydot{}{} phone, \greydot{}{} dispute,\\ \sydot{} home, \yellowdot{} family\end{tabular}                    & \begin{tabular}[c]{@{}l@{}}\bluedot{} \uline{leader}, \greydot{}{} tourism, \greydot{}{} weather,\\ \greydot{}{} restaurant, \greydot{}{} art\end{tabular}    & 3.59            \\ \hline
\multicolumn{2}{l}{G-grand}   & \begin{tabular}[c]{@{}l@{}}\yellowdot{} occupation, \yellowdot{} domestic work\\ \yellowdot{} \uline{white collar job},  \yellowdot{} \uline{feminine}, \yellowdot{} hygiene\end{tabular}
&
\begin{tabular}[c]{@{}l@{}} \bluedot{} pet, \bluedot{} \uline{masculine}, \bluedot{} warmth, \\\sbdot{} eating, \sbdot{} fun \end{tabular}  & \textbf{1.34}                                                           \\\toprule
\multicolumn{2}{l}{\change{Tweets}}     & \begin{tabular}[c]{@{}l@{}}\change{\sydot{} beauty, \yellowdot{} affection, \sydot{} body,} \\ \change{\greydot{}{} fear, \yellowdot{} friends}\end{tabular}  & \begin{tabular}[c]{@{}l@{}}\change{\sbdot{} zest, \sbdot{} joy, \sbdot{} anticipation}, \\ \change{\sbdot{} pride, \greydot{}{} contentment}\end{tabular}  & \change{1.47}              
\\ \bottomrule

\end{tabular}
\caption{Feminine and Masculine Topics for the Birthday Scenario. ``T'' refers to the template dataset and ``G'' refers to the GPT-2 generated dataset.
We use \change{yellow dots \yellowdot{} to denote the most frequent topics (among top 33\%)}, light-yellow dots \sydot{} to denote less frequent topics \change{(from top 33\% to top 66\%)} and grey dots \greydot{}{} for the least frequent topics \change{(after top 66\%)} in the female messages. Similarly, we use the blue color coding for male messages (\change{\bluedot: among top 33\%, \sbdot: from top 33\% to top 66\% and \greydot{}{}: after top 66\% }). \change{We also calculate the odds ratio gap between the most masculine and most feminine words shown in \emph{Gap} column. The higher \emph{Gap} scores are, the more polarized greeting card messages are for women and men on discussed topics.}
}
\label{table:topic_list}
\end{table*}


\begin{table*}[]
\begin{tabular}{l l l l l }
\toprule
\multicolumn{2}{l}{Group}  & Feminine topics                                                                                            & Masculine topics   & Gap                                                                                     \\ \midrule
\multicolumn{2}{l}{T-all}     & \begin{tabular}[c]{@{}l@{}}\yellowdot{} feminine, \yellowdot{} attractive, \sydot{} beauty, \\ \sydot{} affection, \greydot{}{} friends \end{tabular}          & \begin{tabular}[c]{@{}l@{}}\sbdot{} nervousness, \sbdot{} cold, \\ \sbdot{} pain, \sbdot{} body,  \sbdot{} youth\end{tabular} & 1.53 \\ 

\toprule
\multicolumn{2}{l}{G-all}     & \begin{tabular}[c]{@{}l@{}} \yellowdot{} wedding, \yellowdot{} family, \yellowdot{} children, \\ \sydot{} pride, \sydot{} worship \end{tabular}  & \begin{tabular}[c]{@{}l@{}}\sbdot{} leader, \sbdot{} play, \sbdot{} eating, \\ \bluedot{} cooking,  \sbdot{} restaurant \end{tabular} & 2.98               \\ \hline
\multicolumn{2}{l}{G-parents} & \begin{tabular}[c]{@{}l@{}}\yellowdot{} domestic work, \greydot{}{} furniture,\\ \yellowdot{} family, \sydot{} home, \greydot{}{} hipster\end{tabular}                    & \begin{tabular}[c]{@{}l@{}}\greydot{}{} fire, \greydot{}{} art, \greydot{}{} emotional, \\ \sbdot{} writing, \sbdot{} real estate\end{tabular}  & 1.32              \\ \hline
\multicolumn{2}{l}{G-grand}   & \begin{tabular}[c]{@{}l@{}}\yellowdot{}  occupation, \yellowdot{} domestic work, \\\sydot{} dance, \sydot{} messaging, \sydot{} furniture \end{tabular} & \begin{tabular}[c]{@{}l@{}} \sbdot{} masculine, \sbdot{} tourism, \\ \sbdot{} sports,  \greydot{}{} art, \sbdot{} zest\end{tabular}  & \textbf{1.20}                                                          \\ \bottomrule
\end{tabular}
\caption{Feminine and Masculine Topics for the Valentine's Day Scenario. }
\label{table:topic_list_valentines}
\end{table*}

\begin{table*}[]
\begin{tabular}{l l l l l }
\toprule
\multicolumn{2}{l}{Group}  & Feminine topics                                                                                            & Masculine topics & Gap                                                                                     \\ \midrule
\multicolumn{2}{l}{T-all}     & \begin{tabular}[c]{@{}l@{}}\yellowdot{} sadness, \yellowdot{} feminine, \yellowdot{} attractive, \\ \yellowdot{} beauty, \yellowdot{} children \end{tabular}          & \begin{tabular}[c]{@{}l@{}}\bluedot{} wedding, \greydot{}{} childish, \greydot{}{} celebration, \\ \greydot{}{} positive emotion, cheerfulness\end{tabular} & 0.85 \\ 

\toprule
\multicolumn{2}{l}{G-all}     & \begin{tabular}[c]{@{}l@{}} \yellowdot{} feminine, \greydot{}{} fashion, \yellowdot{} childish, \\ \sydot{} appearance, \greydot{}{} fabric \end{tabular}  & \begin{tabular}[c]{@{}l@{}}\bluedot{} leader, \sbdot{} play, \greydot{}{} wedding,\\  \greydot{}{} valuable, \greydot{}{} computer \end{tabular} & 6.41               \\ \hline
\multicolumn{2}{l}{G-parents} & \begin{tabular}[c]{@{}l@{}}\yellowdot{} domestic work, \sydot{} home, \greydot{}{} sports, \\\greydot{}{} music, \greydot{}{} pride\end{tabular}                    & \begin{tabular}[c]{@{}l@{}}\sbdot{} \uline{leader}, \greydot{}{} hipster, \greydot{}{} sexual, \\\greydot{}{} \uline{fashion},\greydot{}{} swearing terms\end{tabular}  & 3.49  \\ \hline
\multicolumn{2}{l}{G-grand}   & \begin{tabular}[c]{@{}l@{}}\yellowdot{} occupation, \yellowdot{} domestic work, \yellowdot{} fabric,\\ \yellowdot{} clothing, \yellowdot{} medical emergence \end{tabular} & \begin{tabular}[c]{@{}l@{}} \bluedot{} journalism, \bluedot{} cold, \bluedot{} music,\\ \bluedot{} sports, \bluedot{} musical\end{tabular}  & \textbf{1.17}                                                          \\ \bottomrule
\end{tabular}
\caption{Feminine and Masculine Topics for the wedding scenario. }
\label{table:topic_list_wedding}
\end{table*}

Table~\ref{table:topic_list}-\ref{table:topic_list_wedding} show the top five masculine and feminine topic lists for the birthday, Valentines' day, and wedding scenario.
In the tables, we use the dark-yellow dots \yellowdot to denote that the most frequent topics (\change{frequency among top 33\%}) in female messages. We use light-yellow dots \sydot to denote less frequent topics (\change{from top 33\% to top 66\%}) and grey dots \greydot for the least frequent topics (\change{after top 66\%}) in the female messages. 
Similarly, we use the blue color coding for male messages (\bluedot: \change{among top 33\%}, \sbdot: \change{from top 33\% to top 66\%} and \greydot: \change{after top 66\%} ).
\change{The \emph{Gap} column represents the polarity of odds ratios between the most masculine and the most feminine topics discussed in Section~\ref{section:method}.} 
Besides the entire dataset (denoted as ``all''), we also split the analysis for different age groups (denoted as ``babies'', ``parents'', and ``grand''.)
We also use \uline{underline} to mark the topics in age groups ({\em e.g.,} feminine in T-babies) if they are repeating the extracted topics in the full corpus ({\em e.g.,} feminine in T-all). 

\textbf{People write more about appearances for \change{women} while more about careers for \change{men}}, as shown in the first row of Table~\ref{table:topic_list}).
We clearly see \emph{``feminine''}, \emph{``attractive''}, and \emph{``beauty''} appearing in all the feminine topic list.
This suggests that people tend to send wishes about appearance regardless of the scenario.
Interestingly, GPT-2 dataset also displays a tendency of generating \emph{``domestic work''} related messages for \change{women}. 
In contrast, the topics for \change{men} are relatively more diverse. 
Still, we observe \emph{``leader''} tend to occur frequently across scenarios (\texttt{T-all} and \texttt{G-all} in Table~\ref{table:topic_list}, \texttt{G-all} in Table~\ref{table:topic_list_valentines} and \ref{table:topic_list_wedding}).
The color of the dots also suggests that these topics frequently occur in the messages.
For example, there are 246 messages for \change{women} while only 112 for \change{men} about the \emph{``appearance''}, and there are 47 messages about the leadership for \change{women} while 134 for \change{men}. 
\change{Similarly, we can also see that people talk more about appearance (e.g., beauty and body) to women in tweets.}

\change{In addition, we analyze topics in greeting card messages to recipients without specific gender indicators in the birthday scenario with our analysis pipeline, by replacing greeting card messages to female recipients and male recipients in the pipeline with the ones to recipients with and without specific gender indicators. We utilize messages in the template dataset with no gender indicators to conduct the study, denoted as ``neutral'' in Table~\ref{table:statistics}. On the other hand, we sample the same number of messages for recipients with specific gender indicators (i.e., messages to female and male recipients combined). The top 5 distinct topics for recipients with specific gender indicators (i.e., binary people) over ones without specific gender indicators  are \emph{``home'', ``royalty'', ``family", ``domestic work'', ``masculine''}, and the top 5 topics for recipients without specific gender indicators compared to binary people are \emph{``work''} (e.g., we pray for numerous years of meaning and \emph{accomplishments} for you.), ``business'' (e.g., If you are half as productive as you are in the \emph{office}, your life is going to be awesome.), ``meeting" (e.g., happy birthday to one of the greatest people I have ever had the pleasure of \emph{meeting}.), ``law'' (e.g., your \emph{discipline} stands amongst the most needed professionals across all niche), ``health'' (e.g.,  we wish you good \emph{health}.), with the \emph{Gap} between most distinctive topics for two groups is 2.95. The topics difference indicates that when people write greeting card messages to binary people, they tend to write more stereotypical topics compared to when they write to recipients without specific gender indicators. 
From the top 5 distinct topics we extract for recipients without specific gender indicators, we can see that most topics are work-orientated and more similar to what we extracted for men, compared to appearance and domestic work-related topics for women.
}
\sherry{(1) so, merging male and female does not make the topical bias cancel out? (2) can you give some more discussions on the results, in comparison to all the man/woman experiments?}

\textbf{Greetings generated by GPT-2 amplifies gender stereotypes.}
\change{The gap for \texttt{G-all} is either higher than or comparable to that for \texttt{T-all}. More specifically, while the differences between the two scores in \emph{Birthday} is trivial, those in \emph{Valentine's day} and  \emph{Wedding} double or even becomes six times larger.}



\textbf{The gender association differs across different age groups.}
The gender distinction is the smallest when people write to the elderly.
While all the non-grandparents age groups have repeating topics from the complete corpus (i.e., ``-all'' rows), the topic overlaps greatly diminishes in the grandparent age group. 
For example, \emph{``healing''}, as the most masculine topic, appears only 3 times for male recipients more than it is for female recipients in the elderly group. 
It means that the gender-associated topics for grandparents are not as strong as for other age groups. 
\change{We further checked if there are potential semantic noises in our text prompt. For instance, our text prompt \texttt{``[scenario prefix] my little baby [girl/boy] [female/male name]!''}  might not refer to literal babies and can be used for one's adult children or one's boyfriend or girlfriend.
We random sampled 50 GPT-2 generated greeting card messages for baby boy and baby girl separately.
Then, we manually evaluate 1) if the generated messages are referring to literal babies or not and 2) if the generated messages are fluent and natural.
Among 100 generated messages, there are 13 invalid ones (6 for baby girls and 7 for baby boys). The high validity rate (i.e., 87\%) shows few semantic noises in our text prompt and generated greeting card messages are of high quality. }

While we were not able to split out or generalize baby-related messages in the other two scenarios, \change{the \texttt{grand} group has the lowest gap score among all age groups, which also holds for the \texttt{Valentine's Day} scenario (Table ~\ref{table:topic_list_valentines}) and the \texttt{Wedding} scenario (Table~\ref{table:topic_list_wedding}). It again shows that the gender distinction is smallest when people write to the elderly compared to the other groups.} 

\sherry{I didn't see an analysis on the twitter data?}




\begin{table}[]
\begin{tabular}{c c c c c}
\toprule
                & GloVe & GloVe Google News & GloVe* \\ \midrule
Birthday        & 0.932        & 0.958       & 0.624  \\ \hline
Valentine's Day & 1.104              &   0.903          & 0.598       \\ \hline
Wedding         &  1.105 & 0.969            & 0.768       \\ \bottomrule
\end{tabular}
\caption{The WEAT scores of feminine/masculine topic lists for different events in the template dataset based on pretrained GloVe~\cite{glove} embedding on different data sources \change{(``GloVe'': original  Glove word embedding; ``GloVe Google News'': pretrained GloVe embedding trained on Google News dataset; ``GloVe*'': GloVe embedding fine-tuned on template dataset).}}
\label{table:weat}
\vspace{-0.5cm}
\end{table}

\textbf{The WEAT score verifies that the topics we found have a clear gender association in word embeddings. }
The WEAT score is in the range of $-2$ to $2$. A high positive score indicates that detected feminine topics are more associated with female attributes in the current embedding space. A high negative score means that detected feminine topics are more associated with male attributes. 
As shown in Table~\ref{table:weat}, we find that \change{WEAT scores across all scenarios and embeddings are all positive, which indicates and verifies the positive association between gender attributes and selected topics.}

\section{\sysname System}

\subsection{Pilot Survey}
\label{section:pilot}

Our analysis shows that greeting card messages to \change{women} lean towards appearance and household while towards career and leadership to \change{men}. AI also exhibits the same characteristics when generating greeting messages. However, whether people are aware of the gender role when greeting and care about having the gender role in their greeting card messages is under-investigated. Here, we want to answer the research questions RQ3 raised in Section~\ref{section:intro}: (\emph{Do people want to be informed of the gender role in their greeting card messages?}) and RQ4 (\emph{If RQ3 stands, how can we help users write greeting card messages? What features do they expect in such a tool?}).

To investigate people's attitudes towards gender roles in greeting card messages, we designed and distributed a pilot survey. \sherry{please don't share this, since it shows the sharer name and breaks anonymity. Just submit all pdf as supplemental material.}
We recruited twenty participants \change{(gender: 6 female, 12 male and two prefer not to self-identify; age: span from 10-20 to 40-50. ethical background: 11 are Asian, 2 are Latino or Hispanic and 6 are White or Caucasian)} \change{by posting volunteer recruitment information on co-authors' social platforms. All recruited participants did not know any content of the project, including any prior findings in our project or our research questions}. 
To set up the context and get a chance to digest the actual messages written by participants --- we first asked them to choose a loved one (or to others) and write a short birthday greeting message.
Afterward, we collected their opinions on showing the gender role in greeting card messages. 
In specific, we asked two questions: (1) whether they would be concerned if their message(s) show a strong gender role, and (2) what features they would expect if they were using a greeting writing tool that informs them about masculine or feminine languages in their written message. We come to some \textbf{C}onclusions derived from our pilot survey:


\textbf{C1: Users would like to be informed about the gender role.}
   Some users \change{(14 out of 20)} are curious about the gender role in their messages, {\em, e.g. .}, ``\textit{I would be curious to know if what I am saying in a greeting does have a strong gender role. I am not sure how the tool would work beyond traditional binary stereotypes, though. I would be concerned about how objective you can make a tool that measures gender orientation in language}'' ($P${16}). 
   In addition, users prefer to know the gender role to avoid offending the receivers of their messages, {\em e.g.}, ``\textit{I would like to be informed to ensure I'm not offending anyone in any sense. Moreover, it will help me understand when I can avoid these remarks}'' ($P${3}).
   
\textbf{C2: However, most users will not be concerned about the gender role.} 
    Some users think greetings are personal and they know their addresses' pronouns, {\em e.g.}, ``\textit{I would say my style of address to someone depends on my individual experiences with that person, rather than any potential stereotypes.}'' ($P${1}).
    They also stated that the inclusion of the gender role is usually intended, {\em e.g.}, ``\textit{I would be concerned. My sister does not necessarily fit stereotypes, and my birthday messages are pretty similar across gender. Adding gender-specific language would likely be different from my usual style}'' ($P${5}). \change{Statistically, 90\% users (18 out of 20) expressed that they would not be concerned about gender roles in their greeting card messages. }

\textbf{C3: Users would like to get some suggestions, but are against the system changing things for them.} 
    \change{Users do not want to be intervened with modifying their greeting messages.}
    For example, $P${4} responded to the second question, saying that ``\textit{I'd be careful about adding a feature that suggests an alternative message because sometimes the person for whom the greeting is meant does not follow traditional gender roles}.''
    $P${3} and $P${16} further emphasized that they do not want to be criticized by the machine. 
    However, they are not against the idea of getting suggestions in two aspects: 
    (1) To diversify their message. $P${8} mentioned ``\textit{I'd love to get some new ideas on what I can include in a greeting message.}''; and (2) to know what contributes to their manifested gender role, o that they can make informed decisions on what to keep or change their messages.

\subsection{Design Requirements}

Based on Section~\ref{section:pilot}, we find that users would like to have an assistant tool to help users write greeting card messages with gender role awareness. Such a tool should satisfy the following requirements:
\begin{itemize}
    \item [$R1$:] \textbf{Bring the gender role awareness to greeting card messages.} More specifically, we need to provide an overall gender perception of a continuous scale (based on users' requirements of going beyond binary gender type) and highlight writings that contribute to the point.
    \item [$R2$:] \textbf{Avoid explicit word attribution and neutralization.} People do not want to be judged, and the appearance of gender roles in greeting card messages is expected.
    \item[$R3$:] \textbf{Associate gender orientation with topics in messages for both analysis and suggestions.} Users want to understand the connection between gender perception and content. They prefer the system to suggest changes but make the modifications themselves. 
\end{itemize}

\begin{figure*}
    \centering
    \includegraphics[width=1.9\columnwidth]{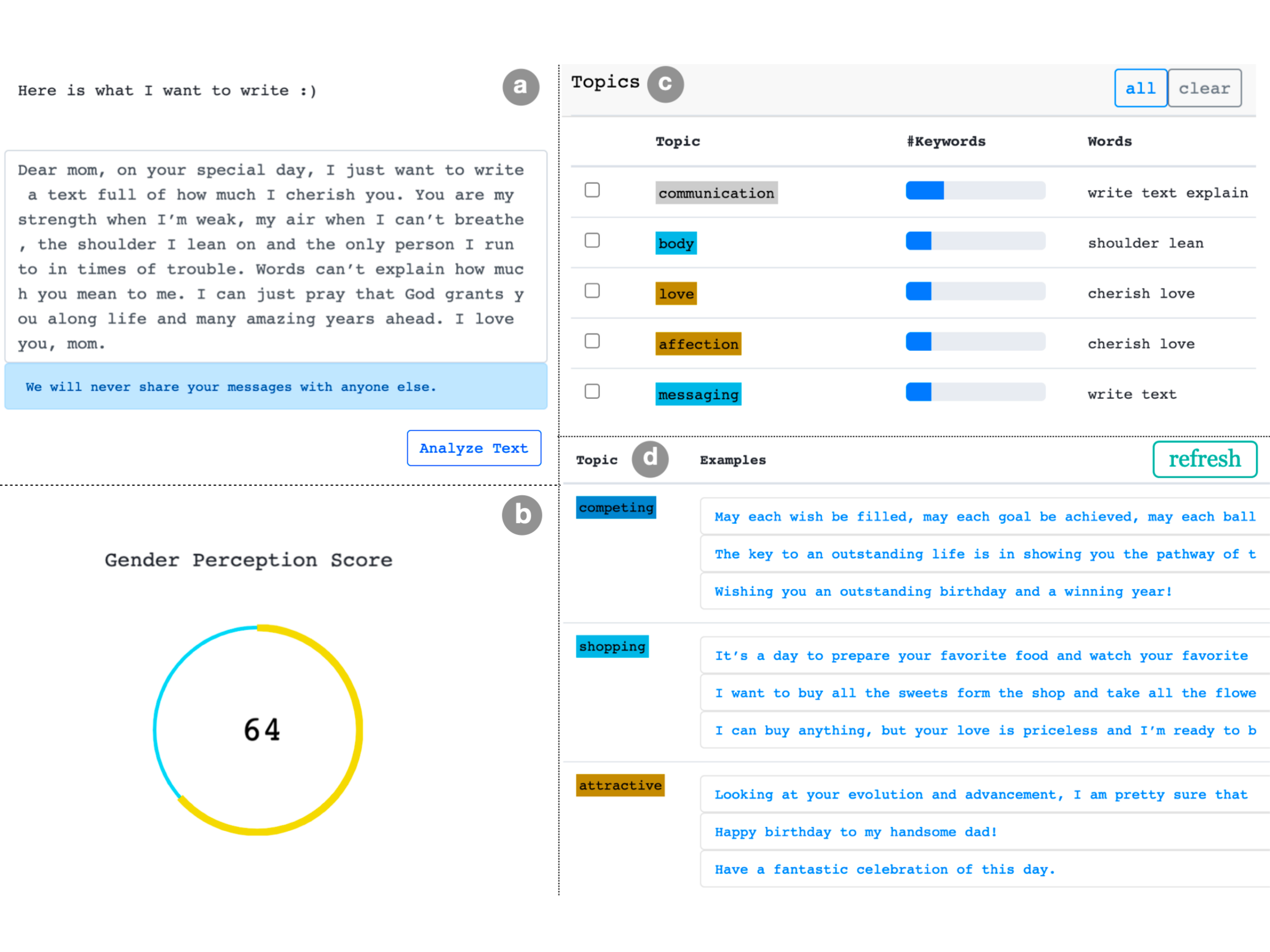}
    \caption{\sysname overview. \sysname contains four panels: \change{message input panel $a$, overall gender perception score panel $b$, topic analysis panel $c$, and topic exploration panel $d$.}}
    \label{fig:overview}
    \vspace{-0.1cm}
\end{figure*}

\change{After digesting users' requirements, we design an interactive \change{analysis} tool, \sysname, to increase users' awareness without intervening in their writings.} \emph{The target of \sysname is to bring gender role awareness to users when writing greeting card messages.} \change{As $P17$ mentioned during the pilot survey, ``You can always just ignore the tool if you want. It is just a warning anyways''. We want to emphasize that there is nothing wrong with having gender roles in greeting card messages. Our design is to increase gender role awareness and help users better understand gender roles in their messages to prevent unconscious gender stereotypes. Users always have full control over their messages and can decide whether to modify greeting card messages on their wishes.}

\subsection{\sysname}

Figure~\ref{fig:overview} shows an overview of \sysname. Users can choose whom they are writing to and write greeting messages in \change{panel $a$}. We use the blue and yellow coding in \sysname to be colorblind-friendly. The color-coding is consistent in the system. Blue, yellow and grey represent more masculine, more feminine and neutral respectively.



\paragraph{Message input panel $a$ and overall gender perception score panel $b$.} Users can write greeting messages in panel $a$. After a user clicks on the analyze text button, \sysname would automatically analyze the user's message and show the analysis result. We provide an overall gender perception score ($R1$), and the lengths of blue and yellow \change{circle fragments represent how much the message leans towards being masculine or feminine. The score inside fragments indicates how much the message leans towards being feminine}.

\paragraph{Topic analysis panel $c$.} \change{We provide top topics that the input message is associated with from the Empath output and sort them based on their occurrence frequencies ($R3$) in panel $c$.} Furthermore, we also list all words falling into each topic to the right of the topic name. In the weight column, the weight bar's length indicates the occurrence frequency of the corresponding topics. When the user clicks on the checkbox in front of each topic, we will highlight the corresponding words related to that topic in the original message with the same color-coding. 

\paragraph{Topic \change{exploration} panel $d$.} In panel $d$, \change{We randomly sampled some topics from the collected template datasets.} Users potentially be interested in adding the topics to the input message. \change{Each displayed topic contains words that fall under the topic based on Empath (e.g., for examples under \emph{competing} topic in Figure~\ref{fig:overview}, there are keywords \emph{achieved, outstanding, winning} in 3 sentences that all fall under \emph{competing} topic based on Empath). All examples we show are retrieved from the collected template datset. If there are more than 3 examples falling under the same topic, we randomly select 3 of them to display in panel $d$. Users can refresh to get another set of topics and corresponding examples if they find the current examples unsatisfactory and want to explore more, and expand the previews to see the complete example content.} We highlight suggested topics with the same color-coding to show the gender role it relates to based on our analysis result. 

\section{Qualitative User Study}
\label{sec:qualititive}
To understand whether \sysname helps people write greeting messages with the gender role awareness,  
\change{we explore two aspects via a qualitative user study:  how much \sysname can help people become more aware of their messages' gender roles ($Q1$), and whether people want to edit their messages after perceiving the awareness ($Q2$).}
To answer these two questions, we conduct a qualitative user study and show our results in this section. 
We have taken standard practice consideration with anonymous data collection and annotations.
This research study has been approved by the Institutional Review Board (IRB) at the researchers’ institution. 

\paragraph{Participants and Apparatus. } We recruited seven student volunteers from the authors' university 
(age 21-27, $\mu = 24$, $\theta^2=3.5$) with a diverse major background, including acting, chemistry, computer science, education, etc., denoted as $P1$-$P7$. All of them have never heard of the project or seen \change{\sysname} before the user study. We make \sysname temporarily publicly available during the study period. \change{Participants used their laptops to conduct the study. It was an online one-on-one study by connecting with participants via video calls. During the study, we encouraged participants to think aloud, share the screen and feel free to stop sharing whenever they type anything personal or do not feel comfortable to share. Please note that we did not save participants' greeting card messages during the study. We also asked for participants' consent if we wanted to quote part of their messages in the paper.} All surveys distributed during the study are available online, and every participant could only submit once.

 
\subsection{Procedure}

\paragraph{Preparation.} Before the study, we asked every user to write birthday greetings consisting of 6-10 sentences to their loved one and recommended them to include specific events. 

\paragraph{Prior Survey.} We asked participants to fill out the prior survey besides basic demographic information of the user and the pronouns of their loved ones.

\paragraph{Using \sysname} We first played a three-minute introduction video of \sysname to show the users how to interact with. We asked them to play with \sysname for fifteen to thirty minutes. During this process, we encouraged the users to think aloud and voice out (e.g., which features are useful, which features are unnecessary, what obstacles they are encountering, etc).

\paragraph{Evaluation Survey.} After playing with \sysname, we asked them to fill out the evaluation survey. The survey contains two parts: \change{1)  To evaluate $Q1$, we first applied the usability metric for user experience (UMUX)~\cite{finstad2010usability}, a four-item Likert scale, to assess participants’ satisfaction with the perceived
usability of \sysname. We then added additional Likert scale items to further measure the level of helpfulness participants perceived with our system. The additional items mainly focus on measuring \emph{whether \sysname is useful for writing greeting card messages, easy to learn, easy to use, helps increase the gender role awareness, and whether they are willing to use it in the future}.} 2) For $Q2$, we use open questions including \emph{Why they changed/kept (a certain part of) the message? What features make you want to change (not change) your message? What do you like about our tool? What do you dislike about our tool? Will you consider using it in the future and why?}

\subsection{Result and Analysis}

\begin{figure*}[t]
        \centering
        \subfigure[Usability] {
                \includegraphics[width=\columnwidth]{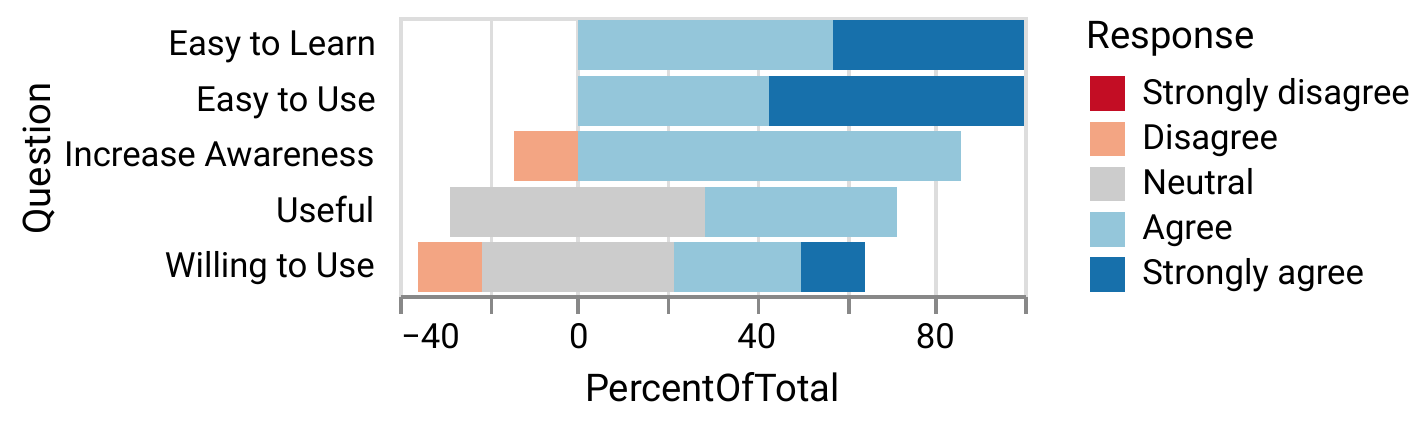}
                \label{fig:usability}
            }
        \subfigure[Most Useful Features]  {
                \includegraphics[width=0.8\columnwidth]{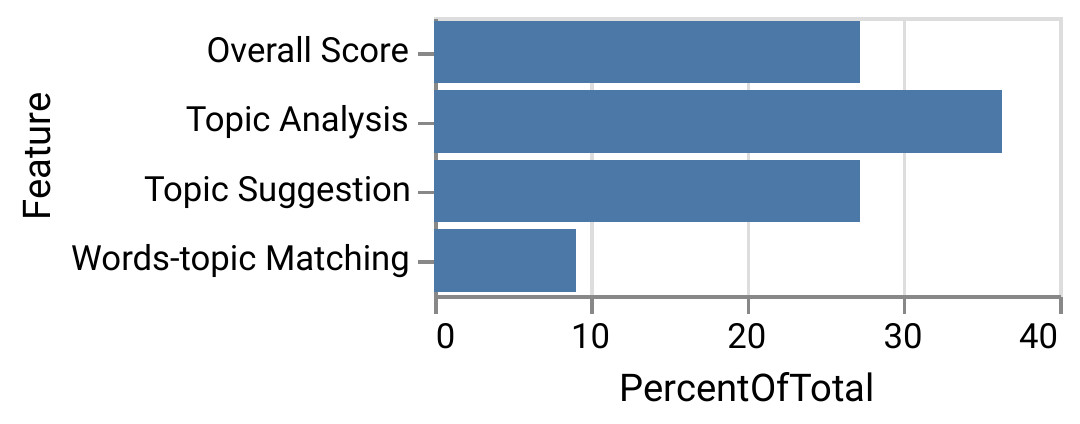}
                \label{fig:features}
        }
        \vspace{-0.15in}
        \caption{\change{The qualitative user study results: including a) diverging stacked bar chart of users' responses on evaluating whether \sysname is useful, easy to learn and use, and increases the gender role awareness. We also asked users if they are willing to use \sysname in the future; b) we collected users' opinions about the most useful features in \sysname. One user can select multiple components as the most useful features; then, we calculate the percentage of votes on one specific feature over total votes for all features.}}
\end{figure*}

We show survey responses to usability questions in Figure~\ref{fig:usability} \change{as a diverging stacked bar chart, where we take ``neutral'' as a baseline so that positive responses are stacked to the right while the negative responses to the left. We describe key insights as follows:}

\textbf{\sysname is easy to learn and use.} All users agreed that \sysname is easy to learn and use.
    
\textbf{\sysname helps increase the gender role awareness ($Q1$).} As shown in Figure~\ref{fig:usability}, six out of seven users agreed that \sysname could help increase the gender role awareness. Some users think \sysname could help decrease gender bias in their messages, {\em e.g.}, ``\textit{If I need to write the greetings to someone I am not very familiar with, like collaborators or colleagues or leaders, I would get the suggestions from the tool to make my message not gender-biased.}'' ($P$1). Other users stated that \sysname could avoid offense to message recipients, {\em e.g.}, ``\textit{As I always watch lots of heroic movies, I am worried that I may write some gender-stereotyped phrasings. I need some tools like this to help me before sending my greeting email to someone else so they will not feel offended or comfortable}'' ($P$6).
    
\textbf{Users may not be willing to change their greeting card messages.} Some users think that if the message recipient is a close friend, they are not concerned about gender bias, {\em e.g.}, ``\textit{since we know each other well and the recipient will not get offended if I use some feminine or masculine words in my wishes.}'' Many users also stated that if the messages contain common experience or hobbies that are gender-biased, it does not matter. $P$6 shared her message to her close male friend with us when she noticed that cooking was associated with domestic work, which is very feminine in our analysis result. She mentioned that she would like to keep it as it is one of the most precious memories between her and her friend, and her friend would not think it was inappropriate. Similarly, P4 showed us part of his message to his female friend ``\textit{No wonder why my game was far more advanced than yours on the court. At least you now have an excuse to be blocked and dunked on.}'', which \sysname \change{extracted as a very} masculine topic ``game'', but $P$4 would not want to change it as they indeed played basketball a lot together.
    
\textbf{People lean to use \sysname in the future ($Q2$).} About half of the users either strongly agreed or agreed that they would use \sysname in the future. $P2$ and $P4$ both indicated that they would \sysname to modify their messages since \sysname helped them \change{identify some words related to negative emotions (e.g., sadness). After consideration, they find the wordings inappropriate and decided to change them} to be more suitable for celebrations. On the other hand, some users also impressed their disagreement, {\em e.g.}, ``\textit{If the message is for someone close like friends or family members, I will write on my own.}'' (P1).

We noticed that $P5$ chose ``disagree'' with both \emph{\sysname could help increase the gender role awareness} and \emph{willing to use}. Reasons $P5$ gave were that ``\textit{I think maybe because the message I wrote was quite neutral, the tool did not point anything out.}'' As we aim to bring the gender role awareness to writing instead of providing editing suggestions, we expect that users like $P5$ who have high gender-role awareness when writing would find \sysname less insightful.



Furthermore, we asked users' opinions on which components they think are most useful. \change{One user can select multiple components as the most useful features; then, we calculate the percentage of votes on one specific feature over total votes for all features in Figure~\ref{fig:features}.} \change{The result shows that users think \emph{topic analysis} is the most useful function, followed by \emph{topic suggestion} and \emph{overall gender role score}.}
For example, $P2$ and $P4$ both found their wordings \change{reveal negative emotions (e.g., sadness) in their greeting card messages and decided to modify their messages}. $P1$ and $P3$ also pointed out that \change{they took some topics in the topic exploration panel $d$} and put them into their messages.

\paragraph{Conclusion of the user study.} 
Although people may not want to change their messages if they know the recipient well, most users found that \sysname helps them increase awareness of the gender role when writing greeting messages, and they are willing to use the system to improve their messages to avoid offending recipients.

\section{Quantitative User Study}

Although over 85\% of users in qualitative study (Section~\ref{sec:qualititive}) self-report that \sysname helps increase the gender role awareness,  we conduct another quantitative study to examine it by setting up three contrast surveys on Amazon Mechanical Turk (MTurk). \change{Ideally, more and more workers can identify the gender role in the greeting card messages and agree with the ground truth labels with different levels of aids from \sysname. The ground-truth labels are created and agreed upon among 3 co-authors. We will explain the details of experiments set up in the following.\sherry{again, annotation criteria?}}

\begin{itemize}

\item \textbf{Survey 1: Unaided.} Given just two greeting messages without the help from \sysname, we ask MTurk annotators to select the one that is more likely to be sent to female recipients and explain the reason why they think it is more feminine.

\item \textbf{Survey 2: Topics.} In addition to the two greeting messages, we provide the topics and corresponding words for each message extracted by \sysname to MTurk workers.

\item \textbf{Survey 3: Topics + Overall Score.} 
In addition to the two greeting messages and the extracted topics, we also provide a Gender Perception Score by \sysname showing how each message leans towards masculine or feminine. The score is higher if the message leans towards the feminine.
\end{itemize}

\subsection{Message Selection Criteria}

Each message has an assigned value in the template dataset showing this gender association where greeting messages with values > 50 are more feminine and < 50 are more masculine. So we divided the corpus into 3 classes with values a) > 51 [feminine], b) < 49 [masculine], c) between 49 and 51 [no gender preference]. We then picked samples based on topics. People tend to talk more about business and leadership to male recipients while talking more about appearance and family-related topics when writing to female recipients. Thus, we picked appearance and family-related topics in the feminine class and business and leadership-related topics in the masculine class. 
Furthermore, we asked \change{3} co-authors to pair-wisely pick more feminine messages and check whether the results match the result \sysname provides, which we use as the ground truth for later. \change{The gender perception scores from \sysname are all consistent with ground-truth labels created by co-authors.}
In the end, we picked 20 pairs of greeting messages, including 3 types: female v.s. male, female v.s. neutral and neutral v.s. male. 

\subsection{Procedure}

We recruited 25 high-quality (approval rate: >=98\%, HITs: >=1000) MTurk workers located in the United States for each survey. 
Each MTurk worker in each survey (Unaided, Topic, or Topic + Overall Score) is assigned to read 20 pairs of greeting messages with/without additional information from \sysname and select the more feminine ones and explain the reasons. \change{Each pair took an average of half a minute to finish complete with compensation of \$0.15 ( 20 pairs will take 10 minutes with a compensation of \$3)}, above the US Federal minimum wage. 
An annotator can participate only once in one survey and \change{will be} filtered out as a qualification for the other two surveys.
In our instruction, we clarified that our surveys are used to evaluate people's perception of gender distinction in the greeting messages and describe the information we provide in the surveys along with a sample example. 

\subsection{Result}

\begin{table*}[]
\begin{tabular}{l l l l l}
\toprule
\textbf{Codes}                           & \textbf{Explanations and Examples} &  \textbf{\text{Unaided}} &  \textbf{\text{Topics}} & \textbf{\text{Topics+score}}
\\

\toprule

Words 
& \begin{tabular}[c]{@{}l@{}}  Pick on certain words and phrasings \\ \emph{{\color{gray}``everyday hero is more masculine."}}\end{tabular}  &   133 (73.9\%)  & 130 (72.2\%) &  146 (81.1\%)  \\

Topics & 
\begin{tabular}[c]{@{}l@{}}  Based on conveyed topics  \\ \emph{{\color{gray}``it speaks to the person's nurturing qualities''}}\end{tabular}
&    26 (14.4\%)    & 85 (47.2\%)   & 48 (26.7\%)     \\

Overall&
\begin{tabular}[c]{@{}l@{}}  Perceived the overall gender role \\ 
\emph{{\color{gray}``This sounds like a message to a female sibling''}} \end{tabular} 
&  29 (16.1\%)   &   22 (12.2\%)    & 52 (28.9\%)        
\\ 
Experience & 
\begin{tabular}[c]{@{}l@{}} Based on their experience or gender stereotype \\
\emph{{\color{gray}``less likely for men to be thanked for cleaning''}} \end{tabular} 
&    35  (19.4\%)
& 25 (13.9\%) & 20 (11.1\%)


\\ \bottomrule
\end{tabular}
\caption{The codebook of how annotators choose the more feminine message, with the number of self-reports.}
\label{table:codebook}
\end{table*}



\begin{figure*}
    \centering
    \includegraphics[width=1.4\columnwidth]{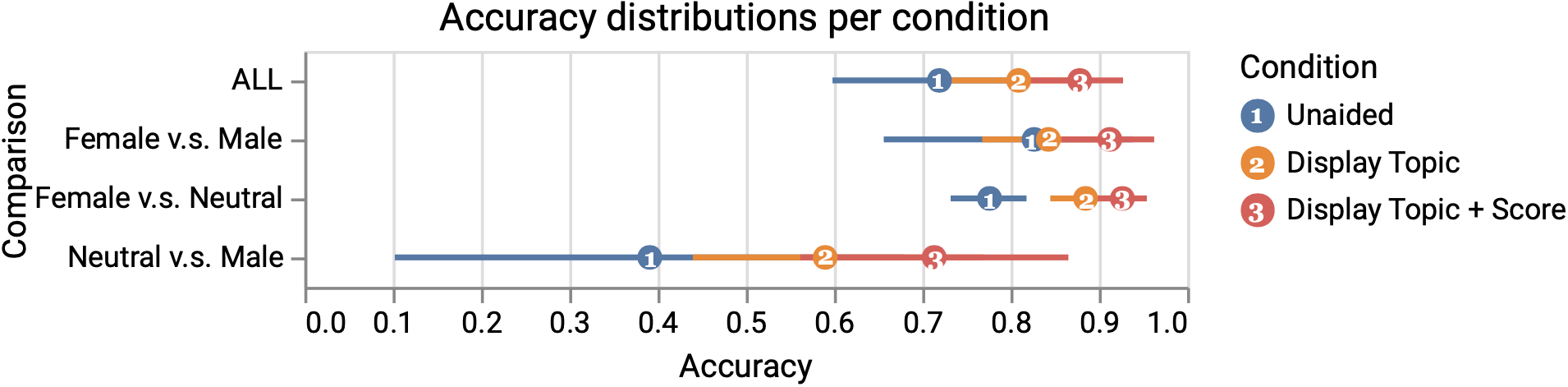}
    \caption{\change{The accuracy distribution of question pairs in 3 surveys, where dots are the average accuracy, and lines crossing the dots represent each condition's confidence interval. The average accuracy across all comparisons for three surveys is consistently improved for all combinations. It indicates the effectiveness of \sysname on improving gender role awareness.}}
    \label{fig:accuracy}
\end{figure*}

\textbf{\sysname helps to increase gender role awareness in general.} We compared results from the three surveys. As shown in \autoref{fig:accuracy}, the average accuracies across all comparisons for three surveys are 70.1\%, 80.0\%, and 87.7\%, respectively.
The result indicates that the correctness of identifying gender perception got improved after introducing \sysname.
The \emph{topic+score} condition yielded the highest accuracy, which may not be too surprising, given the summary score essentially reveals the gender association.

\textbf{\change{\sysname offers useful information for workers to identify gender roles in greeting card messages.}}
To better understand whether workers' rationales differ across conditions, we coded their self-reported explanations for the ranking.
\change{Three of the coauthors independently read a subset of the responses to identify emergent codes and created a codebook (Table~\ref{table:codebook}) using a discussion period.
Using this codebook, for each survey, we coded a sample of 180 random worker responses (out of 500): 160 were unique, and 20 overlapped between annotators, allowing us to compute inter-annotator agreement and manually annotate what workers rely on to make decisions based on the codebook.}
We achieved reasonably high agreements (Cohen's kappa score~\cite{cohen1960coefficient} $\kappa=0.76$).
We observed that, as expected, annotators in \emph{Topic} and \emph{Topic+score} relied more on the additional information provided by \sysname.
For example, those in the \emph{Topics} condition relied more heavily on the topics (with almost 50\% annotators), compared to the \emph{Unaided} group (with 15\%), and \emph{Topics+score} displays an increment in \texttt{Overall} (which includes the of reliance on the score.)
Curiously, the reliance on topics dropped in \emph{Topics+score}; Instead, annotators seem to rely more heavily on keywords in this case.
We suspect that annotators would search for keywords that supply the ranking given by the score.

\textbf{People have a clearer mental model on female associated topics.} 
We further separate the evaluation for different types of questions. 
Figure~\ref{fig:accuracy} shows that although comparisons that involve female-oriented messages yielded reasonable accuracies, the accuracy of male vs. neutral pairs was considerably harder, regardless of the condition.
This observation is also aligned with our results in Table~\ref{table:topic_list}-\ref{table:topic_list_wedding}, i.e., masculine topics show more diversity than feminine topics.
\change{It might result from 1) people may have a more fixed impression on how to greet \change{women}, and 2) it is inherently harder to choose a more ``feminine'' message when there are only ``neutral'' and ``masculine'' messages.}

\textbf{Culture factors and experience influence how people perceive gender role \change{\cite{Vijver2007CulturalAG, Soltanpanah2018ACE}}}, \change{which we also observe in our study for greeting card messages.} 
When inspecting explanations, we also noticed interesting contrasts that are likely due to different cultural backgrounds and experiences. 
For example, for the message ``I propose a toast to you, darling friend, and our incredible friendship. You are aging like a fine wine, and it is an honor to be a part of your life.''
While one annotator perceived ``aging like fine wine'' as offensive and unwanted for \change{women} (and think it should be sent to males), another put ``darling and fine wine are more often used by \change{women}' girl friends'' (and think it should be sent to \change{women}). 
Future research is needed to consider cultural aspects for analysis.

\textbf{It is important to avoid gender stereotypes in greeting card messages.} Explanations that annotators wrote during our quantitative study further prove the importance of preventing gender stereotypes in greeting card messages.
About 15\% annotators expressed that they identified gender association based on their own experience or gender stereotypes.  For example, ``\textit{Domestic work and cleaning typically refer to a female's role at home. Women are generally the ones who are expected to do the household chores and take care of the family.}''. Besides, about 5\% of participants also explicitly criticized the message, {\em, e.g.,} ``\textit{Most women are infantilized due to being in a male-dominated society. Women are not taken seriously as leaders.}''; ``\textit{Many women are expected to have so many responsibilities and make sure they look good. Especially women who have children to raise. They are often overworked and do not get much recognition for all of the things they do.}''; ``\textit{Most women have to be multitaskers and multitalented in ways that are not expected of men.
An example is the expectation for a woman to be a good mom, a good wife, never need to take a break, and still find time to keep herself looking attractive. A female recipient should be seen as someone to admire when she can do all things and make it look effortless.}'' This collected feedback further reinforces the importance of our work, and how \sysname could be used to advocate a more fair and diverse environment.





\section{Discussion}

\subsection{Implication}
Our work provides a quantitative understanding of the gender roles in greeting card messages by analyzing both the template dataset of greeting messages and the AI-generated (GPT-2) message dataset. Greetings to \change{women} lean towards their appearances and household, and messages to \change{men} lean towards their career achievement. We design and develop \sysname to help visualize such gender related information, and and provide relevant example greetings as references for users. 
\change{Both our qualitative and quantitative user studies demonstrated the effectiveness and usability of \sysname.} 
We hope that \sysname could bring more insights into how to avoid potential gender stereotypes when users write greeting template messages.
We envision that \sysname can have big impacts on multiple aspects to different stakeholders.
Individuals, companies or institutes need to send greetings and wishes to their friends, families, employers or partnership on various scenarios. We hope that \sysname could also help them better polish their messages and avoid potential unconscious stereotypes to advocate and encourage a more fair and diverse environment. 


From a generalization perspective, \sysname can be useful in many scenarios. \change{The topic analysis and overall gender perception analysis can be extended to analyze and understand the gender perception in the general text.} For example, our system can be adapted to benefit a broad set of gender bias analysis tasks from Wikipedia pages, blogs, Twitter, etc. Researchers who work on the natural language generation tasks can also use \sysname to check if their results may include potential gender stereotypes. \change{It will contribute to preventing unwanted bias before we put AI into real-world production.} Moreover, our analysis pipeline and GreetA are general and can be applied in other applications beyond checking greeting card messages, such as game design, helping and raising awareness about imposter syndrome and checking algorithmic bias. 

Note that \sysname might be used for malicious purposes. We have the topic exploration feature in \sysname to assist users in coming up with potential topics to talk about in their greeting card messages. The source is the template dataset that we collected from the greeting card websites. As our analysis shows, the greeting card messages from the collected corpus may contain potential stereotypes. \change{Thus, \sysname might serve as a tool that highlights stereotyped messages if users consistently refresh topics and pick messages with potential gender stereotypes on purpose.}

Our analysis shows that GPT-2 generated messages amplify the bias in human-written messages.  People should be more careful when considering AI in real-world applications. AI researchers should also put more effort into mitigating the bias from different levels ({\em e.g.}, corpus level for our case) and avoiding biased models.


\subsection{Limitation and Future work}
\label{section:limitation}

This work is subject to several limitations. First, we applied Empath~\cite{Fast2016EmpathUT} to analyze topics; although it covers various topics, its measures mainly rely on individual lexicons and do not consider the contextual information. For example, Empath would extract topic \emph{weather} for message \emph{``You are a cool person''} based on linguistic information. However, here \emph{``cool''} indicates the personality rather than the weather. Besides, \change{Empath is bad at dealing with figurative speeches. For example, Empath would extract topic \emph{body} for message \emph{ ``You are the shoulder I lean on''}, but here shoulder is a figurative saying of emotional support. To quantitatively assess the performance of Empath, we randomly select 50 greeting card messages, put selected messages to \sysname, and we evaluate how many among top 5 topics Empath suggested are natural and correct to humans. As a result, 228 out of 250 (i.e., 91.2\%) word-topic suggestions are correct. It again shows the efficiency of using Empath in our task, but also shows the room for improvement.}  
\change{Future work can utilize human annotation or other topic extraction techniques to identify fine-grained topics and better deal with figurative speech.} 

Secondly, we \change{mainly} collect and use the template dataset that people perceive as ``ideal'' messages to analyze how people write greeting card messages. They might be different from what people write in the real world, as they might be more formal and contain the less personal experience. However, the real greeting card messages are hard to get as they are inherently private, and people may not want to share with others. \change{Although we collect 500 real greeting card messages from Twitter, how large these messages can represent is unsure. Future work that collects and utilizes a larger-scale of greeting card messages in social platforms (e.g., Twitter and Reddit) could be complementary to our work. } 
It is also important to understand and analyze greeting card messages \change{to non-binary people, missing from our work because of the scarce source available online. Instead, our work utilizes collected greeting card messages with no specific gender indicators on the birthday scenario and understands the topic difference between birthday greeting messages for binary people and recipients without specific gender indicators. Future research can build upon our work, incorporate more experiments on analyzing greeting card messages to non-binary people across various scenarios and provide assistance when greeting non-binary people. We hope that our work can contribute to building a more inclusive and diverse environment.}



\section{Conclusion}

In this work, we collected a large-scale greeting card messages corpus from eight popular greeting suggestion websites and generated an artificial dataset using the natural language model GPT-2. Via a set of thorough analyses, we found that the topics in greeting card messages are indeed gender oriented and greetings to \change{women} lean towards their appearance, and \change{men} lean towards their career achievement. Our pilot survey showed that people want to be aware of gender role in their greeting messages. In response, we designed and developed \sysname based on a deep understanding of users' requirements. \sysname visualizes the gender orientation in greeting messages and their topical aspects, as well as recommends potential topics for writing better greetings.  The qualitative and quantitative  studies showed that \sysname effectively brings the gender role awareness to users. Our work is the first to quantitatively analyze the gender role using statistical NLP tools in greeting messages. \sysname is also the first interactive visualization system that helps bring gender role awareness to people when writing greeting messages. 
\section*{acknowledgement}
This work was supported in part by grant from by the Russell Sage Foundation. Diyi Yang was supported by Microsoft Research Faculty Fellowship. 
The authors thank Nan Xu from USC for helpful discussions and anonymous reviewers for constructive feedback and comments to improve the draft.

\bibliographystyle{ACM-Reference-Format}
\bibliography{ref}

\appendix

\appendix




\end{document}